\documentclass[11pt,english]{scrartcl}
\usepackage{lmodern}
\usepackage[T1]{fontenc}
\usepackage[latin9]{inputenc}
\usepackage{geometry}
\usepackage{amsmath}
\usepackage{mathtools}
\DeclarePairedDelimiter{\ceil}{\lceil}{\rceil}

\usepackage{amssymb}
\usepackage{amsfonts}
\usepackage{graphicx}
\usepackage{color}
\usepackage{soul} 
\usepackage{comment}
\usepackage{amsthm}
\usepackage{bbm}
\usepackage{cleveref}
\usepackage{mathtools}

\newtheorem{theorem}{Theorem}
\newtheorem{corollary}{Corollary}

\newtheorem{lemma}{Lemma}
\newtheorem{proposition}{Proposition}
\newtheorem{claim}{Claim}
\newtheorem{remark}{Remark}
\newcommand{\dfn}{\stackrel{\triangle}{=}}
\newcommand {\exe} {\stackrel{\cdot} {=}}
\newcommand {\gexe} {\stackrel{\cdot} {\ge}}
\newcommand {\lexe} {\stackrel{\cdot} {\le}}

\newcommand {\bs} {\mbox{\boldmath $s$}}

\newcommand {\bx} {\mbox{\boldmath $x$}}
\newcommand{\ux}{\underline{x}}
\newcommand{\uhx}{\underline{\hat{x}}}
\newcommand{\ug}{\underline{g}}
\newcommand{\uhg}{\underline{\hat{g}}}
\newcommand {\by} {\mbox{\boldmath $y$}}
\newcommand {\bz} {\mbox{\boldmath $z$}}

\newcommand {\bE} {\mbox{\boldmath $E$}}

\newcommand {\tP} {\tilde{P}}
\newcommand {\Dm} {D^-}
\newcommand {\hH} {\hat{H}}

\newcommand {\bX} {\mbox{\boldmath $X$}}
\newcommand {\bY} {\mbox{\boldmath $Y$}}

\newcommand{\calC}{{\cal C}}

\newcommand{\calE}{{\cal E}}
\newcommand{\calF}{{\cal F}}
\newcommand{\calG}{{\cal G}}
\newcommand{\calH}{{\cal H}}
\newcommand{\calI}{{\cal I}}

\newcommand{\calP}{{\cal P}}

\newcommand{\calS}{{\cal S}}
\newcommand{\calT}{{\cal T}}
\newcommand{\calU}{{\cal U}}

\newcommand{\calX}{{\cal X}}
\newcommand{\calY}{{\cal Y}}
\newcommand{\calZ}{{\cal Z}}

\allowdisplaybreaks
\title{Universal Randomized Guessing Subjected to Distortion}

\author{Asaf Cohen\footnote{A.\ Cohen is with the School of Electrical and Computer Engineering, Ben-Gurion University of the Negev, Beer Sheva 8410501, Israel ( email: coasaf@bgu.ac.il).} \hspace{1pt} and Neri Merhav\footnote{N.\ Merhav is with Andrew and Erna Viterbi Faculty of Electrical and Computer Engineering, Technion-Israel Institute of Technology (IIT), Haifa 3200003, Israel (e-mail: merhav@ee.technion.ac.il).}}

\begin{document}
\maketitle
\begin{abstract}
In this paper, we consider the problem of guessing a sequence subject to a distortion constraint. Specifically, we assume the following game between Alice and Bob: Alice has a sequence $\bx$ of length $n$. Bob wishes to guess $\bx$, yet he is satisfied with finding any sequence $\hat{\bx}$ which is within a given distortion $D$ from $\bx$. Thus, he successively submits queries to Alice, until receiving an affirmative answer, stating that his guess was within the required distortion.

Finding guessing strategies which minimize the number of guesses (the \emph{guesswork}), and analyzing its properties (e.g., its $\rho$--th moment) has several applications in information security, source and channel coding. Guessing subject to a distortion constraint is especially useful when considering contemporary biometrically--secured systems, where the ``password" which protects the data is not a single, fixed vector  but rather a \emph{ball of feature vectors} centered at some $\bx$, and any feature vector within the ball results in acceptance.

We formally define the guessing problem under distortion in \emph{four different setups}: memoryless sources, guessing through a noisy channel, sources with memory and individual sequences. We suggest a randomized guessing strategy which is asymptotically optimal for all setups and is \emph{five--fold universal}, as it is independent of the source statistics, the channel, the moment to be optimized, the distortion measure and the distortion level. 
\end{abstract}


\section{Introduction}\label{sec. Intro}
Consider the problem of guessing a realization of a random $n$--vector $\bX$ over an alphabet $\calX$ subject to a distortion measure $d$. Specifically, assume Alice has such a realization $\bx$. Bob wishes to guess the value of $\bx$, yet is content with discovering any $n$--vector $\hat{\bx}$ over $\hat{\calX}$ which is within a distortion $D$ away from $\bx$. He thus submits to Alice a sequence of \emph{queries} of the form ``Is $\hat{\bx}_i$ close to $\bx$ within distortion $D$?", $i=1,2, \ldots$, until receiving a positive answer. Let $G(\bx)$ denote the \emph{guesswork}, namely, the smallest $i$ for which $\hat{\bx}_i$ is close enough to $\bx$, in the sense of achieving an average distortion smaller than or equal to $D$ under the measure $d:\calX \times \hat{\calX} \to \mathbb{R}^+$. Clearly, Bob is interested in devising a strategy which minimizes $G(\bx)$ in some sense. If $\bx$ is a realization of a random vector, Bob may wish to minimize the expected value of $G(\bX)$ or some other moment of it. However, in many cases, Bob will have, for example, only limited knowledge (if any) on the distribution of $\bX$ or the exact distortion required for success. Moreover, $\bx$ might be a single, fixed individual sequence, without any underlying probability distribution. Bob might also have limited resources, preventing him from remembering which queries were already submitted, or, alternatively, it might be the case where multiple ``Bobs" try to guess $\bx$ simultaneously, without any coordination. To make the matter even worse, Bob's queries might also reach Alice indirectly, e.g., through a noisy channel. Thus, a guessing strategy which is universal and efficient in the senses mentioned above is highly desirable. In other words, we seek a strategy which is independent of the source distribution, the memory structure, the moment to be optimized, the distortion measure and the distortion value, the channel between Bob and Alice, and, finally, a strategy which requires no memory of previously submitted queries and can be applied by multiple guessers simultaneously, yet, despite all the above, asymptotically achieves the optimal guessing performance.   

Guesswork problems have numerous applications in information theory and several other fields. Coding applications were first given in \cite{wozencraft1957sequential,Arikan96,PfisterSullivan} by Wozencraft, Arikan and Pfister and Sullivan, respectively. In \cite{AM98joint}, Arikan and Merhav used a guessing decoder for joint source--channel coding. Fixed-to-variable source coding without a prefix constraint, or \emph{one-shot coding}, were considered by several groups in \cite{Szpankowski11F2V,Kontoyiannis14optimal_lossless,Courtade_Verdu_14,Kosut_universal_F2V_17}. In fact, recently Kumar \emph{et al.}\ \cite{kumar2020guessing} nicely tied several source coding, guessing and task partitioning problems \cite{Bunte2014Tasks} to the same minimization problem, resulting in simple proofs for several known results in the area. Guessing can also be seen as a method to quantify the complexity of a rate--distortion encoder, that is, the number of guesses until a vector \emph{within distortion $D$} of the source vector is found corresponds to the number of candidate codebook vectors the encoder examines until it finds an adequate codeword \cite{AM98}. Another important application for guessing is directly guessing the noise over a channel for capacity-achieving decoding. Duffy \emph{et al.}\ \cite{Duffy2019Capacity} suggested that the receiver can simply order noise sequences from most likely to least likely, then subtract these noise sequences from the received signal. The first noise sequence subtracted which results in a codeword corresponds to ML decoding. We review additional information--theoretic results in \Cref{related work}.

Recent trends in information security raised further attention to guesswork problems, adopting it as a proxy to password strength \cite{bishop1995improving,dell2010password,kelley2012guess,komanduri2011passwords}. A detailed discussion is given in \cite[Section I]{MerhavCohen20}, supporting the applicability of guesswork analysis, and its insights, to practical problems. It is important to note, though, that when focusing on such applications, one cannot limit the analysis to a simple, i.i.d.\ case with a known distribution, as sequences used in information security applications tend to have memory \cite{malone2012investigating,1341406,Vishwakarma14dictionary} and their exact distribution is rarely known \cite{malone2012investigating}.  More importantly, due to the many difficulties in managing passwords, contemporary authentication systems use different kinds of biometric measures \cite{8590812}. In such systems, a \emph{feature vector} is extracted from the biometric data (e.g., fingerprint, ECG, iris photo or even a sequence of keystroke timings) and fed to a previously--trained unit (e.g., a neural network) which either accepts or rejects the data. In what follows, we argue that not only such a feature vector might have a complex and unknown distribution, any system which either accepts or rejects it \emph{must bare some level of distortion between the trained data and the tested data} (unlike ordinary passwords), as such feature vectors are extracted from real--world, human metrics and behaviours. 

In \cite{page2015ECG}, Page \emph{et al.}\ suggested ECG-based biometric authentication. The raw QRS--complex\footnote{These are the typical downward--upward--downward deflections usually seen on an ECG signal \cite{wiki_QRS}.} of the signal was classified using a neural network, with only minor filtering as pre--processing. Raw ECG signals were used in \cite{hammad2019ECG} as 2D images. Again a (convolution) neural network was used for classification. ECG signals where used for authentication in \cite{aziz2019ecg} as well. This time, several features, such as variance, skewness and bandwidth occupied were extracted, and a Support Vector Machine (SVM) was used for classification. SVM was also used for voice biometrics in \cite{Boles2017VoiceBiometrics}. Additional techniques for voice authentication are discussed in \cite{meng2020active}, including both traditional MAP-based techniques as well as modern Deep Neural Networks. Detailed surveys can also be found in \cite{mahfouz2017survey,8590812}. The underlying assumption in all is that \emph{slight variations in the input data to the authentication system should not result in a different decision}. Again, this should be contrasted with passwords, which require a perfect match for acceptance.

In fact, the classification literature is focused on constructing robust neural networks and SVMs, such that slightly distorted input would not distort the classification. Indeed, slightly distorted inputs at selected places which cause the classifier to output different classes are known in the literature as \emph{adversarial examples}, and quite a few techniques were suggested to combat them (and output the same class as the non--distorted input), including adding noise \cite{gu2014towards,carlini2017provably}. Bi and Zhang gave motivating examples in \cite{bi2005support}, referring to classifying sentences from speech recognition, or classifying images after an image processing phase. In both, it is clear that the initial recognition or processing might introduce errors to the input, and multiple instances of the input will result in multiple, slightly different, feature vectors, \emph{yet all should be mapped to the same class} (or have the same accept or reject result). Hence, the classifier should be robust to such differences (distortions, in the context of this work), in the sense of producing the same class as the non-distorted input. Moreover, the noise model considered in \cite{bi2005support} is i.i.d.\ with zero mean, which fits the distortion model we consider in this paper. A deterministic formulation is given in \cite{Tsai2021AdversarialRobustness} by Tsai \emph{et al.}, where one wishes the output vector of the neural network for a perturbed input to be close to the original output, if the perturbed input vector is in a ball around the original one. In \cite{xu2017feature}, Xu \emph{et al.}\ suggested \emph{feature squeezing} of the input data before learning or classification. For images, the technique reduces to bit depth reduction, which essentially means, again, that a ``ball" of different, yet close, input vectors (images) is treated as one for both learning and classification, hence, in the context of guessing, \emph{any guess within this ball should be successful}.
\subsection{Main Contributions}
In this paper, we suggest randomized, five-fold universal guessing strategies for guessing subject to a distortion measure. Specifically, we start with memoryless sources, and give a randomized, universal guessing distribution which achieves the optimal guesswork exponent for any memoryless source distribution, any moment (of the guesswork) and any distortion measure and level. As guessing sums up to drawing sequences independently from a given universal distribution, this method is inherently distributed, allowing multiple and asynchronous guessers to guess independently, yet asymptotically achieve the minimal number of queries. We continue to the problem of guessing through a noisy channel. In this case, again, the guesser (Bob) wishes to toss a sequence which is within a given distortion from the true word, yet now guesses pass through a noisy (memoryless) channel before arriving at Alice. We show that now the universal guessing distribution is not only universal in the distribution of the source, the moment and the distortion, but is also universal in the channel transition probability matrix. We also intuitively motivate the resulting guesswork exponent, and show how it implicitly includes the optimal, non-universal guessing distribution in an alternative expression.

We then revisit the problem for sources with memory. We give a guessing distribution, based on the Lempel--Ziv 78 (LZ78, \cite{ZL78}) parsing rule, which is shown to be universal for any $\Psi$-mixing source\footnote{A wide family of sources with memory, which includes Markov models, Unifilar sources, etc. Roughly speaking, these are sources which might have long-term dependencies, yet the dependency between any two blocks decays with the distance between the blocks. A formal definition is given in \Cref{Guessing for Sources with Memory}.} and any moment or distortion level. For the direct part, we first formally define a block--memoryless approximation for such sources. We then derive the exponent achieved by the universal guessing distribution, and conclude with a matching converse.

Last but not least, we turn to the fourth setup, in which the sequence to be guessed is an individual sequence, without any underlying probability distribution. We show that the LZ-based guessing distribution discussed above is asymptotically optimal, in the sense of achieving the best possible guesswork exponent compared to any \emph{finite--state guessing machine}.

Since the distribution suggested for memoryless sources can be viewed as a special case of the distribution for sources with memory, the results of the above four setups reveal that \emph{a single guessing distribution, is, in fact, asymptotically optimal for all setups, and hence universal in a multitude of dimensions}: the source distribution and memory, the distortion measure and level and the memoryless channel the guesses might pass through. Moreover, this distribution has a simple and efficient implementation, which, in short, requires only applying an LZ decoder on a sequence of i.i.d.\ uniform bits. Finally, we also note that the guesswork exponents under this distribution meet very general converse bounds, which do not assume randomized guessing and, in fact, allow for a wide variety of guessing strategies.  

The rest of this paper is organized as follows. In \Cref{related work}, we thoroughly discuss related work. \Cref{sec. statement} includes the required notation and the problem statement. \Cref{sec. main} includes the main results. \Cref{sec. three lemmas} includes three important lemmas, which stand at the basis of the proofs. Indeed, the achievability result for memoryless sources easily follows from the first lemma, and matches a known lower bound. The achievability result for noisy guessing also follows directly, this time from the second lemma in \Cref{sec. three lemmas}. Its converse, however, is new and appears separately in \Cref{sec. converse for noisy}. For sources with memory, both the achievability and the converse are more involved, hence their complete proofs appear in \Cref{sec. memory}. Finally, the appendix includes additional proofs of some technical lemmas and claims for the first three setups, as well as the proofs for the achievability and converse parts of the fourth setup, in which Bob tries to guess an individual sequence.  
\section{Related Work}\label{related work}
Guesswork was information--theoretically formalized by Massey \cite{Massey94}. Later, Arikan \cite{Arikan96} computed the exponential rate of guesswork for memoryless sources. Specifically, it is was proved to be the R\`enyi entropy of order $\frac{1}{2}$. Arikan and Merhav \cite{AM98} were then the first to study guesswork under a distortion constraint, and devised a strategy which is
universal and asymptotically optimal for memoryless sources, both in the source distribution and the moment of the guesswork. Recently, randomized guessing subject to distortion was studied by Kuzuoka in \cite{kuzuoka2020asynchronous,kuzuoka2021asynchronous}. It was shown that for memoryless sources, the optimal guesswork moment can be achieved by randomized guessing. Yet, the distribution used was the tilted distribution similar to \cite{SHBCM19}, thus depending on both the source distribution and the moment $\rho$. Hence, for memoryless sources, the strategy in \cite{AM98} is independent of the distribution and the moment, yet depends on the distortion measure and value, while the tilted distribution in \cite{kuzuoka2020asynchronous,kuzuoka2021asynchronous} is independent of the distortion measure and value, yet depends on the source distribution and moment. 

\emph{Ordering and Universality.} Indeed, since the strategy suggested in \cite{AM98} is to sort the sequences to be guessed in an increasing order of the \emph{empirical rate--distortion function}, the method is independent of the true source distribution and the moment order. This universal strategy extends the strategy for guessing without distortion, which sorts the sequences in an increasing order of their empirical entropy. Such orderings were, in fact, found useful in other applications, e.g., Weinberger \emph{et al.}\ \cite{Weinberger_universal_ordering_92} ordered strings by the size of their type-class before assigning them codewords in a fixed-to-variable source coding scheme, Kosut and Sankar used it in \cite{Kosut_universal_F2V_17} and Beirami \emph{et al.}\ \cite{Beirami15geometric} showed that the dominating type in guesswork (the position of a given string in the list) is the largest among all types whose elements are more likely than the given string. Note that in all these cases, sequences are sorted, then guessed one after the other, using, e.g., a list of sequences. This process is inherently centralized\footnote{From a practical viewpoint, this process applies in cases where a pre-compiled list of usernames and passwords is used - ``hard-coding" the guessing strategy \cite{owens2008study,tirado2018new}. This should be compared to distributed brute--force attacks \cite{SHBCM19,SHBCM20}, and the asynchronous, randomized strategies we suggest herein.}. However, all methods hint towards the structure of \emph{universal guessing distributions, which assign probabilities to sequences in an analogous manner}. Such distributions played a key role in \cite{MerhavCohen20} and will play a role here as well. To conclude the discussion on universality, Sundaresan \cite{sundaresan2007guessing} also considered guessing under source uncertainty. The redundancy as a function of the radius of the family of possible distributions was defined and quantified. As expected (\cite{AM98}), for discrete memoryless sources this redundancy tends to zero as the length of the sequence grows without bound. 

\emph{Sources with Memory.} Guesswork for Markov processes was first studied by Malone and Sullivan \cite{malone2004guesswork}, and later extended by Pfister and Sullivan in \cite{PfisterSullivan}. Hanawal and Sundaresan proposed a large deviations approach \cite{5673955}, generalizing results from \cite{Arikan96} and \cite{malone2004guesswork}. In \cite{christiansen2013guesswork}, again via large deviations, Christiansen \emph{et al.}\ proposed an approximation to the \emph{distribution} of the guesswork. Recently, Merhav and Cohen \cite{MerhavCohen20} considered universal randomized guessing for sources with memory, and showed that an LZ78 \cite{ZL78} parsing procedure can be utilized to construct such a universal, asymptotically optimal distribution. Motivated by these results, Merhav \cite{Merhav20a} showed that, in fact, an LZ78 decoder fed by purely random bits is asymptotically optimal, not only for sources with memory, but also for \emph{individual sequences}, compared to any other finite--state machine. The above works, however, considered only the lossless case, i.e., with no distortion allowed.

The need to account for sources with memory in guessing problems was also verified experimentally. Dell'Amico \emph{et al.}\ \cite{dell2010password} evaluated the probability of guessing passwords using dictionary-based, grammar-free and Markov chain strategies, using existing data sets of passwords. Guessing strategies which account for memory performed better. Moreover, the need to tune memory parameters was clear, thus highlighted the urgency for universal strategies. Bonneau \cite{bonneau2012science} also acknowledged the difficulty of coping with passwords from unknown distributions. Needless to say, biometric authentication methods mentioned in \Cref{sec. Intro} also stress out the fact that the sequences to be guessed usually have an unknown distribution, and might contain highly dependent measurements \cite{page2015ECG,hammad2019ECG,Boles2017VoiceBiometrics}. 

\emph{Noisy Channels.} Guessing over a noisy channel was studied by Christiansen \emph{et al.}\ in \cite{christiansen2013guessing}. Therein, the sequence to be guessed was assumed i.i.d., but \emph{was passed through a channel} before reaching the guesser. Thus, if the channel introduces erasures, the goal is to fill the missing gaps. In a sense, this model was extended by Salamatian \emph{et al.}\ in \cite{SHBCM20}, where \emph{multiple guessers} tried to fill the missing gaps, or correct errors introduced by the channel, using either a centralized or a decentralized approach. On the other hand, \emph{submitting guesses through a noisy channel} was considered by Merhav in \cite{Merhav20b}.

\emph{A Multitude of Models.} The current literature on guesswork is growing, with numerous new results on various aspects of the problem. Christiansen \emph{et al.}\ \cite{christiansen2015multi} considered a multi-user case, where an adversary has to guess $U$ out of $V$ strings. Beirami \emph{et al.}\ \cite{7282958} further defined the inscrutability of a source as the number of guesses for the above problem, and the inscrutability rate as its exponential growth rate. The same paper also showed that ordering strings by their type-size in ascending order is a universal guessing strategy. Yet, again, both \cite{christiansen2015multi} and \cite{7282958} considered a single attacker, with the ability to create a list of strings and guess one after the other. Non--asymptotic results which can be applied to guesswork were given in \cite{courtade14} by Courtade and Verd\'u. In \cite{sason2018tight}, Sason derived bounds on the R\'{e}nyi entropy of a function of a random variable, and applied them to derive non-asymptotic bounds on the difference between the exponent in guessing the original random variable, and that of guessing the (possibly non one--to--one) function. Several works considered guessing with side information \cite{Arikan96,4036408,christiansen2013guessing,SHBCM20,Sason_Verdu_18}. Yona and Diggavi \cite{yona17bias} considered the problem of guessing a word which has a similar \emph{hash function} as the source word - a highly practical scenario as passwords are rarely stored as is, and only a hash is used. Ardimanov \emph{et al.}\ \cite{Ardimanov20Oracle} considered the problem of guessing with the use of an oracle, which the guesser can use before the guessing game begins. On the same line of work, Weinberger and Shayevitz \cite{weinberger2020guessing} considered the problem where the guesser can receive information from a helper which saw the true word through a memoryless channel.
\section{Notation and Problem Statement}\label{sec. statement}
\subsection{Basic Notation}
Throughout, random variables will be denoted by capital
letters, their realizations will be denoted by the
corresponding lower case letters, and their alphabets
will be denoted by calligraphic letters. Random
vectors and their realizations will be denoted in the bold. For example, the random vector $\bX=(X_1,\ldots,X_n)$ may take a specific vector value $\bx=(x_1,\ldots,x_n)$
in $\calX^n$, the $n$--th order Cartesian power of $\calX$, which is
the alphabet of each component of this vector. Sources will be denoted by the letters $P$ or $Q$. The expectation operator will be denoted by $\bE\{\cdot\}$, and $\calI\{\cdot\}$ will denote the indicator function. The entropy of a distribution $Q$ on $\calX$ will be denoted by $H_Q(X)$ where $X$ designates a random variable drawn by $Q$. $D(Q\|P)$ will denote the relative entropy between $Q$ and $P$. The rate--distortion function of a memoryless source $Q$ at distortion level $D$ will be denoted by $R(D,Q)$. 

$\exp_2\{v\}$ denotes $2^v$ and $\log$ denotes $\log_2$. For two positive sequences $a_n$ and $b_n$, $a_n\exe b_n$ will
stand for equality on the exponential scale, that is,
$\lim_{n\to\infty}\frac{1}{n}\log \frac{a_n}{b_n}=0$. Similarly,
$a_n\lexe b_n$ means that
$\limsup_{n\to\infty}\frac{1}{n}\log \frac{a_n}{b_n}\le 0$, and so on.
When both sequences depend on a vector, $\bx\in\calX^n$, namely, 
$a_n=a_n(\bx)$ and $b_n=b_n(\bx)$,
the notation $a_n(\bx)\exe b_n(\bx)$ means that the asymptotic convergence is
uniform, namely, 
\begin{equation}\label{def. exe}
\lim_{n\to\infty}\max_{\bx\in\calX^n}\bigg|\frac{1}{n}\log
\frac{a_n(\bx)}{b_n(\bx)}\bigg|=0.
\end{equation}

The empirical distribution of a sequence $\bx\in\calX^n$, which will be
denoted by $\hat{P}_{\bx}$, is the vector of relative frequencies
$\hat{P}_{\bx}(x)$
of each symbol $x\in\calX$ in $\bx$. Information measures associated with empirical distributions will be denoted with `hats' and subscripted by the sequences from which they are induced. For example, the entropy associated with $\hat{P}_{\bx}$, which is the empirical entropy of $\bx$, will be denoted by $\hat{H}_{\bx}(X)$. Analogously, to stress out a distribution, e.g., $Q$, under which an information measure is calculated, we use subscripts as well, e.g., $H_Q(X)$. With a slight abuse of notation, when clear from the context, we use similar notation for multi-variate measures, e.g., $I_Q(Y;Y')$ denotes the mutual information under the joint distribution $Q_{YY'} = Q_Y Q_{Y'|Y}$, which will be abbreviated by $Q$. Moreover, when the $Y$--marginal of such a distribution $Q$ will be the empirical distribution of a specific $\by$, we will use the notation $Q_Y$ and not $\hat{P}_{\by}$, if the specific $\by$ is clear from the context.  

The \emph{type class} of $\bx\in\calX^n$, i.e., the set of all vectors in $\calX^n$ with empirical distribution $\hat{P}_{\bx}$, will be denoted by $\calT(\bx)$. Similarly, $\calT(Q)$ will denote the type class of a specific (empirical) distribution $Q$. E.g., $\calT(Q_{X|Y}|\by)$ will denote the set vectors $\bx\in\calX^n$ whose relative frequencies equal $Q_{X|Y}$ for the specific $\by$.

\subsection{Guessing Subject to a Distortion Measure}
Alice selects a random $n$--vector $\bX$,
drawn from a finite alphabet source $P$. Bob, which is unaware of the
realization of $\bX$, submits a sequence of guesses in the form of yes/no queries:
``Is $\bX$ close within distortion $D$ to $\hat{\bx}_1$?'', ``Is $\bX$ close within distortion $D$ to $\hat{\bx}_2$?'', and so on, where $\hat{\bx}_i \in \hat{\calX}^n$ for each $i$, until receiving a positive answer\footnote{Similar to the common definition in rate--distortion theory, $\hat{\calX}$, the reconstruction alphabet, may be different from the source alphabet $\calX$. Clearly, the scenario in which $\hat{\calX}=\calX$ is a special case.}. Specifically, Alice gives a positive answer iff Bob's guess is \emph{within an average distortion $D$} of Alice's realization, that is 
\begin{equation}\label{distortion measure}
\frac{1}{n}\sum_{i=1}^n d(x_i,\hat{x}_i) \leq D,
\end{equation}
where $d:\calX \times \hat{\calX} \to \mathbb{R}^+$ is a single--letter distortion measure. With a slight abuse of notation, we may write $d(\bx,\hat{\bx})$ for the $n$--letter distortion, that is, $\sum_{i=1}^n d(x_i,\hat{x}_i)$. Given $\bx\in\calX^n$, let $\calS(\bx)=\{\hat{\bx}:~d(\bx,\hat{\bx})\le nD\}$. Clearly, for a given realization $\bx$, Bob's goal is to guess a sequence $\hat{\bx} \in \calS(\bx)$.

When considering the deterministic setting, a \emph{guessing list}, $\calG_n$, is an ordered list of all members of $\hat{\calX}^n$,
that is, $\calG=\{\hat{\bx}_1,\hat{\bx}_2,\ldots,\hat{\bx}_{|\hat{\calX}|^n}\}$, and it is associated with a \emph{guessing function}, $G(\bx)$, which is the function that maps $\calX^n$ onto
$\{1,2,\ldots,|\hat{\calX}|^n\}$ by assigning to each $\bx\in\calX^n$ the smallest integer $i$ for
which $\hat{\bx}_i$ satisfies the distortion constraint \eqref{distortion measure}. Namely, $G(\bx)$ is the
number of guesses required until success, using $\calG_n$, when $\bX=\bx$. In this setting, the goal is to devise a guessing list $\calG_n$ that minimizes
a certain moment of $G(\bX)$, namely, $\bE\{G^\rho(\bX)\}$, where $\rho > 0$ is a given positive real.

In the randomized setting, on which we focus in this work, the guesser simply sequentially submits a sequence of random guesses, each one drawn independently according to a certain probability distribution $\tilde{P}(\hat{\bx})$. This setting consumes less memory (compared to deterministic guessing which stores the list $\calG_n$) and needs no synchronization. Note that, in general, $\tilde{P}(\hat{\bx})$ may depend on $i$, where $i-1$ is the number of guesses submitted thus far. However, such a dependence \emph{will require} memory and synchronization between multiple guessers. More importantly, we will later see that a distribution which is independent of $i$ \emph{still achieves the converse bounds for all setups considered} in this paper, hence there is no significant gain in considering a sequence of guessing distributions, indexed by the number of guesses thus far. 

Clearly, in this setting, the goal is to devise a \emph{guessing distribution $\tilde{P}$} that minimizes
a certain moment of the guesswork. However, we will also be interested in devising a distribution which is independent of the source Alice has, the moment of the guesswork to be minimized, the distortion measure $d(\cdot,\cdot)$ and the distortion level $D$. 

For a given $\bx$, the expected number of guesses until a sequence with a small enough distortion is guessed is independent of the underlying distribution used to draw $\bx$, namely, $\bE\{G^\rho|\bx\}$ depends solely on the distribution used to guess, $\tP(\hat{\bx})$, and the set $\calS(\bx)$. In fact, for $\bE\{G^\rho|\bx\}$ we have the following result, which directly utilizes \cite[Lemma 1]{MerhavCohen20}.
\begin{lemma}\label{geometric lemma}
\begin{equation}
	\bE\{G^\rho|\bx\}=\sum_{k=1}^\infty k^\rho \tP[\calS(\bx)](1-\tP[\calS(\bx)])^{k-1}
	\exe \frac{1}{(\tP[\calS(\bx)])^\rho}.
\end{equation}
\end{lemma}
The expectation in \Cref{geometric lemma} is with respect to the guessing distribution, yet for a given source sequence $\bx$. When $\bx$ is random, the source distribution has to be taken into account. We can now clearly define the main goal of this paper: devise universal guessing distributions, which achieve the smallest possible \emph{guesswork exponent}. That is, we will be interested in the limit $\lim_{n \to \infty} \frac{1}{n}\log \bE\{G^\rho\}$, where the expectation in $\bE\{G^\rho\}$ is over both the source distribution and any randomization in the guessing strategy. Specifically, we will devise universal distributions, compute their resulting exponent $\bE\{G^\rho\}$, and prove matching converse results, showing that no better exponent can be achieved.
\subsection{Universal Guessing Distributions}\label{guessing distributions}
We will utilize two important universal distributions and later show that these distributions, used for the respective source models, will suffice to achieve the universality goals stated above. A key step in the process will be to bound \emph{the probability of the set $\calS(\bx)$ under these distributions}. This is done in \Cref{sec. three lemmas}.

Specifically, for memoryless sources, we consider a sequence of randomized guesses, all drawn independently from the
universal distribution,
\begin{equation}\label{universal distribution for memoryless}
\tilde{P}(\hat{\bx})=\frac{\exp_2\{-n\hat{H}_{\hat{\bx}}(\hat{X})\}}{\sum_{\hat{\bz}\in\hat{\calX}^n}\exp_2\{-n\hat{H}_{\hat{\bz}}(\hat{Z})\}},
\end{equation}
for all $\hat{\bx}\in\hat{\calX}^n$. Note that this distribution satisfies $\tP(\hat{\bx})\exe 2^{-n\hat{H}_{\hat{\bx}}(\hat{X})}$, and was used in \cite{MerhavCohen20} for $D=0$.

For sources with memory, we suggest a guessing distribution based on the incremental parsing procedure of LZ78 similar to \cite{MerhavCohen20}. The incremental parsing procedure of \cite{ZL78} (see also \cite[Subsection 13.4.2]{CT06}) is a sequential procedure for parsing a sequence, such that each new parsed phrase is the shortest string that has not been obtained before as a phrase. For example, the binary string 
\begin{equation}
\hat{\bx} = 01101101110010111
\end{equation}
is parsed as 
\begin{equation}
0,1,10,11,01,110,010,111.
\end{equation}
For a string $\hat{\bx}$, let $c(\hat{\bx})$ denote the number of such distinct phrases. In the example above, $c(\hat{\bx})=8$. In general, for a binary sequence of length $n$, $c(\hat{\bx}) \leq n/[(1-\epsilon_n) \log n]$, where $\epsilon_n \to 0$ as $n \to \infty$ (\cite[Theorem 2]{LZ76}). Roughly speaking, when coding $\hat{\bx}$ losslessly, one may simply describe, for each new phrase, the location of its previously appearing prefix, and the new symbol. This results in a codelength of about $c(\hat{\bx}) \log c(\hat{\bx})$ bits. Specifically, let $LZ(\hat{\bx})$ be the \emph{LZ code length} (in bits) of the sequence $\hat{\bx}= \hat{x}_1, \hat{x}_2, \ldots, \hat{x}_{n}$. We define a probability distribution $\tP$ over $\hat{\calX}^n$ as follows
\begin{equation}\label{tP with LZ}
\tP(\hat{\bx}) = \frac{2^{-LZ\left(\hat{\bx}\right)}}{\sum_{\hat{\bz} \in \hat{\calX}^n}{2^{-LZ\left(\hat{\bz}\right)}}}.
\end{equation}
Note that since LZ is a uniquely decodable code, we have $\sum_{\hat{\bx} \in \hat{\calX}^n}{2^{-LZ\left(\hat{\bx}\right)}} \leq 1$ and hence $\tP(\hat{\bx}) \ge 2^{-LZ\left(\hat{\bx}\right)}$.
\section{Main Results}\label{sec. main}
In this section, we summarize the main results in this paper. We begin with memoryless sources. Then, we provide the results for guessing through a noisy channel, followed by the results for sources with memory. Finally, we conclude with results for guessing an individual sequence.
\subsection{Memoryless Sources}
\begin{theorem}\label{Direct for memoryless}
For any memoryless source $P$ over $\calX$, if sequences are selected independently at random according to the distribution $\tP(\hat{\bx})\exe 2^{-n\hH_{\hat{\bx}}(\hat{X})}$, then, the $\rho$-th guesswork moment, subject to a distortion constraint $D$, satisfies
\begin{equation}\label{guessing exponent memoryless}
    \bE\{G^\rho\} \exe \exp_2\left\{n\max_Q[\rho R(D,Q)-D(Q\|P)]\right\}.
\end{equation}
\end{theorem}
The proof of \Cref{Direct for memoryless} follows from \Cref{First Lemma on PS} and appears right after it, in \Cref{sec. three lemmas}. 

Note that \Cref{Direct for memoryless} meets the exponential order of the lower bound in \cite{AM98}. Hence, guessing using the distribution in \Cref{Direct for memoryless} is asymptotically optimal, and we have the following.
\begin{corollary}\label{coro. memoryless}
For any memoryless source $P$ over $\calX$, the best achievable $\rho$-th guesswork moment, subject to a distortion constraint $D$, is achievable by randomized guessing and we have
\begin{equation}
   \lim_{n \to \infty} \frac{1}{n}\log \bE\{G^\rho\} = \max_Q[\rho R(D,Q)-D(Q\|P)].
\end{equation}
\end{corollary}
Clearly, the exponent in \Cref{coro. memoryless} is similar to that in \cite{AM98}, yet it can be achieved in a randomized manner, with no memory (to hold a list) or synchronization.
\subsection{Noisy Guessing for Memoryless Sources}
We consider the following scenario, studied first in \cite{Merhav20b} for $D=0$. Alice draws a random $n$--vector,
$\bY=(Y_1,\ldots,Y_n)$, from a discrete memoryless source
(DMS), $P$, of a finite alphabet, $\calY$.
Bob, who is unaware of the realization of $\bY$, sequentially submits to Alice
a (possibly, infinite) sequence of guesses, $\bx_1,\bx_2,\ldots$, where each $\bx_i$ is a vector
of length $n$, whose components take on values in a finite alphabet, $\calX$.
Before arriving to Alice, each guess, $\bx_i$, undergoes a discrete memoryless channel (DMC), defined by
a matrix of single--letter input--output transition probabilities,
$W=\{W(y|x),~x\in\calX,~y\in\calY\}$. Let $\bY_1,\bY_2,\ldots$ be the
corresponding noisy versions of $\bx_1,\bx_2,\ldots$, after being corrupted by
the DMC, $W$. Alice sequentially examines the noisy guesses and she returns
to Bob an affirmative feedback upon the first match within distortion $D$, that is, $d(\bY,\bY_i)\le nD$.
Clearly, the
number of guesses, $G$, until the first successful guess, is a random
variable that depends on the source vector $\bY$ and the guesses,
$\bY_1,\bY_2,\ldots.$. It is given by
\begin{equation}
G=G(\bY,\bY_1,\bY_2,\ldots)=\sum_{k=1}^\infty
	k\cdot\calI\{d(\bY,\bY_k)\le nD\}\cdot\prod_{i=1}^{k-1}[1-\calI\{d(\bY,\bY_i)\le nD\}].
\end{equation}
For a given list of guesses, $\calG_n=\{\bx_1,\bx_2,\ldots\}$,
$\bx_i\in\calX^n$, $i=1,2,\ldots$, the $\rho$--th moment of $G$ is given by
\begin{equation}
\label{G}
\bE_{\calG_n}\{G^\rho\}=\sum_{\by\in\calY^n}P(\by)\cdot\sum_{k=1}^\infty k^\rho\cdot
	W(\calS(\by)|\bx_k)\cdot\prod_{i=1}^{k-1}[1-W(\calS(\by)|\bx_i)],
\end{equation}
where 
\begin{equation}
W(\by|\bx)=\prod_{t=1}^nW(y_t|x_t)
\end{equation}
and $\calS(\by)=\{\by':~d(\by,\by')\le nD\}$.
Randomized guessing lists (where the deterministic guesses, $\{\bx_i\}$, are replaced by
random ones, $\{\bX_i\}$) are allowed as well. In this case, eq.\ (\ref{G})
would include also an expectation w.r.t.\ the randomness of the guesses.

Let $P$, $W$, and $\rho\ge 0$ be given.
For two given distributions, $Q_X$ and $Q_Y$, defined on $\calX$ and $\calY$,
respectively, define
\begin{equation}
\Gamma(Q_X,Q_Y)=\inf\{D(\tilde{Q}_{Y|X}\|W|Q_X):~(Q_X\odot\tilde{Q}_{Y|X})_Y=Q_Y\},
\end{equation}
where 
\[
D(Q_{Y|X} \| P_{Y|X} | Q_X) = \sum_{x\in \calX}Q_X(x)\sum_{y\in \calY} Q_{Y|X}(y|x) \log \frac{Q_{Y|X}(y|x)}{P_{Y|X}(y|x)}
\]
and the notation $(Q_X\odot\tilde{Q}_{Y|X})_Y=Q_Y$ means that the $Y$--marginal
induced by the given $Q_X$ and by $\tilde{Q}_{Y|X}$ is constrained to be the given $Q_Y$,
i.e., $\sum_{x\in\calX}Q_X(x)\tilde{Q}_{Y|X}(y|x)=Q_Y(y)$ for all $y\in\calY$.
Next, define
\begin{equation}
\Gamma(Q_Y)=\inf_{Q_X}\Gamma(Q_X,Q_Y)=\inf_{Q_{X|Y}}D(Q_{Y|X}\|W|Q_X).
\end{equation}
Finally, we define
\begin{equation}\label{1st}
	E_W(\rho)=\sup_{Q_Y}\bigg\{\rho\hat{R}_W(D,Q_Y)
	-D(Q_Y\|P)\bigg\},
\end{equation}
where
\begin{equation}\label{def. hatR}
	\hat{R}_W(D,Q_Y)\dfn\inf_{\{Q_{Y'|Y}:~\bE_Qd(Y,Y')\le D\}}
        [I_Q(Y;Y')+\Gamma(Q_{Y'})].
\end{equation}
We argue that $E_W(\rho)$ is the best achievable guessing exponent.
\begin{theorem}\label{theorem noisy}
Assume Alice draws a sequence of length $n$ from a  memoryless source $P$ over $\calY$, and Bob's guesses, before arriving to Alice, undergo a discrete memoryless channel (DMC) $W=\{W(y|x),~x\in\calX,~y\in\calY\}$. Then, the optimal guesswork $\rho$-th moment exponent at distortion level $D$ is achievable using a randomized guessing strategy, which is universal in $P, W, d(\cdot,\cdot), D$ and $\rho$, and satisfies
\begin{equation}
    \lim_{n \to \infty}\frac{1}{n}\log \bE\{G^\rho\} = E_W(\rho).
\end{equation}
\end{theorem}
The proof of the direct part of \Cref{theorem noisy} follows from \Cref{Second Lemma on PS} and appears right after it, in \Cref{sec. three lemmas}. The proof of the converse part is given in \Cref{sec. converse for noisy}.

Similar to \eqref{guessing exponent memoryless}, for the clean memoryless case, the exponent in \eqref{1st} has a clear structure of a \emph{rate--distortion} function, minus the divergence between the distribution for which the function is computed, and the original distribution. The rate--distortion function, however, is not simply the minimal value of the mutual information subject to the distortion constraint, but, rather, includes a correction term, $\Gamma$, to reflect the penalty the guesser suffers due to the noisy channel. In fact, this is the same correction term as in \cite{Merhav20b}, when no distortion is allowed.

The exponent in \eqref{1st} has two alternative forms, which facilitate numerical evaluation and give an operational meaning. 
\begin{lemma}\label{lemma for alternative form}
The function $E_W(\rho)$ in \eqref{1st} has the following two alternative forms,
\begin{eqnarray}\label{alternative}
	E_W(\rho)&=&\sup_{s\ge 0}\inf_V\left\{\log\left(\sum_y\frac{P(y)}
	{\left[\sum_xV(x)U_s(x,y)\right]^\rho}\right)-\rho sD\right\}\label{alternative form 1}\\
	&=&\sup_{s\ge 0}\inf_{M\in\calC\calH(W)}\left\{\log\left(\sum_y\frac{P(y)}
	{\left[\sum_{y'}M(y')e^{s[D-d(y,y')]}\right]^\rho}\right)\right\},\label{alternative form 2}
\end{eqnarray}
where the infimum on $V$ is over all probability distributions on $\calX$, $U_s(x,y)\dfn\sum_{y'}W(y'|x)e^{-sd(y,y')}$, $M(y')\dfn\sum_xV(x)W(y'|x)$ and 
$\calC\calH(W)$ is the convex hull of $\{W(\cdot|x),~x\in\calX\}$.
\end{lemma}
\Cref{lemma for alternative form} is proved in \Cref{proof for alternative form}. Note that the form \eqref{alternative form 1} can be easily evaluated for small alphabets and simple channels (e.g., binary alphabets, Hamming distortion and the binary symmetric channel). The form \eqref{alternative form 2} has an operational meaning. To see this, first recall the result in \cite[eq. (19)]{SHBCM19}, stating that $\bE\{V_\rho(X)\} = \sum_{x\in \calX} \frac{P(x)}{\hat{P}(x)^\rho}$, where $V_\rho(X) = {G(X)+\rho-1 \choose \rho}$ and $n \choose k$ is the generalized binomial coefficient (\cite[eq. (11)-(12)]{SHBCM19}). As $\bE\{V_\rho(X)\}$ approximates the $\rho$--th moment of the guesswork up to a constant factor,  $\sum_{x\in \calX}\frac{P(x)}{\hat{P}(x)^\rho}$ also has the interpretation of a guesswork's $\rho$--th moment, when the source distribution is $P(x)$ and the guessing is randomized according to $\hat{P}(x)$. A similar interpretation was given in \cite[Section V.2]{Merhav20b}. In the problem at hand, $P(y)$, at the numerator of \eqref{alternative form 2}, is the source distribution. To see that the bracketed expression at the denominator (without the power of $\rho$) can be viewed as the \emph{success probability of a single guess under some guessing strategy}, note that $M(y')$ is the channel output distribution when its input is i.i.d.\ according to $V$. Thus the bracketed expression at the denominator can be thought of as the Chernoff bound for the probability of a single successful guess within distortion $D$. To conclude this discussion, note that the minimizing $V$ has the operative meaning of the optimal (non--universal) 
i.i.d.\ random guessing distribution. 
\begin{remark}[A sufficient condition for no--noise--penalty] 
It is interesting to identify the scenario where there is no loss due to the channel $W$. If the test channel, $Q_{Y'|Y}^*$, that achieves the ordinary
rate--distortion function, $R(D,Q_Y)$, happens to induce a marginal distribution $Q_{Y'}$ such that $\Gamma(Q_{Y'})=0$ for any $Q_Y$ (or, at least, for the dominant one), then the guessing exponent 
is the same as in the clean--channel case.
\end{remark}
\subsection{Sources with Memory}\label{Guessing for Sources with Memory}
Our model for sources with memory will be that of $\Psi$-mixing sources. Consider a stationary source $\{X_i\}_{i=-\infty}^{i=\infty}$ over a probability space $(\Omega,\calF,P)$. Let $\calF_j^l$ denote the $\sigma$-field of events generated by the random variables $\{X_m\}_{j \leq m \leq l}$. A \emph{measure of dependence}, $\Psi(k)$, for the source $\{X\}$ is defined by 
\begin{equation}
\Psi(k) = \sup_{A \in \calF_{-\infty}^{0}, B\in \calF_{k}^{\infty}, P(A)>0, P(B)>0} \left| \frac{P(A\cap B)}{P(A)P(B)}-1\right|.
\end{equation} 
A stationary source is said to be \emph{$\Psi$-mixing} if $\Psi(k) \to 0$ as $k \to \infty$.

Before presenting the main result for mixing sources, a few additional definitions are in order. For any positive integer $n$, denote by $R_n(D,P^n)$ the rate-distortion function of a \emph{block-memoryless} source of block size $n$ and a distribution $P^n$ over $\calX^n$, at a \emph{per--letter} distortion level $D$ (that is, $nD$ per--block). Note that we do not normalize the rate by $n$, hence $R_n(D,P^n)$ represents the number of bits required per block of size $n$. Now, for any stationary source $P$ with a marginal distribution $P^n$ define the guessing exponent at distortion level $D$ and moment $\rho$ as
\begin{equation}\label{def. exponent}
E_P(D,\rho) = \liminf_{n\to\infty}\sup_{Q^n}\frac{1}{n} \left[\rho R_n(D,Q^n) - D\left(Q^n||P^n\right)\right].
\end{equation}
\begin{remark}\label{continuity} 
For each $n$, $\sup_{Q^n}\frac{1}{n}R_n(D,Q^n)$ is monotonic in $D$ and hence continuous everywhere with the possible exception of countably many points. Thus, $\liminf_{n\to\infty}\sup_{Q^n}\frac{1}{n}R_n(D,Q^n)$ has the same property. 
\end{remark}
The main result for sources with memory is the following theorem.
\begin{theorem}\label{main theorem memory}
For any stationary $\Psi$-mixing source $P$, the optimal guesswork $\rho$-th moment exponent at distortion level $D$ is achievable using a randomized guessing strategy, which is universal in $P, d(\cdot,\cdot), D$ and $\rho$, and satisfies
\begin{equation}
\limsup_{n\to\infty}\frac{1}{n}\log\bE\{G^\rho\} = E_P(D,\rho).
\end{equation}
\end{theorem}
The randomized guessing strategy is based on the incremental parsing procedure of LZ78. It was described in \Cref{guessing distributions}. For the proof of \Cref{main theorem memory}, \Cref{subsec. lemma for memory} includes a key result, focusing on the probability this distribution assigns to sequences satisfying the distortion constraint. Then, using this result, \Cref{parsing section} completes the details of the direct. \Cref{section converse} includes the matching converse.
\subsection{Guessing an Individual Sequence}
In \cite{Merhav20a}, the guessing game was extended to \emph{guessing an individual sequence}, that is, a single, deterministic sequence without any underlying probability model. In such a setup, it is unrealistic to measure performance against any guessing strategy, and one tries to compete against any \emph{finite--state guessing machine}. This is, in fact, the common practice in many other setups involving individual sequences, including compression, gambling, prediction, denoising and encryption \cite{ZL78,Feder91Gambling,FMG92,Weissman2001Twofold,Modha2004FS-RD,DUDE2005,Merhav12Encryption}.

A finite--state guessing machine is defined by the tuple $(\calU,\hat{\calX},\calZ, \Delta,f,g)$, where $\calU$ is its input alphabet, assumed binary, $\hat{\calX}$ is the output alphabet, $\calZ$ is a finite set of states, with $|\calZ| = s$, $\Delta: \calZ \to \{0,1,2,\ldots\}$ defines the number of input bits used at each state, and $f,g$ are the output and next--state functions, that is, $f: \calZ \times \calU^* \to \hat{\calX}$ and $g: \calZ \times \calU^* \to \calZ$. When a binary sequence $u_1,u_2,\ldots$, $u_i \in \calU$, is fed to the guessing machine, the output is a sequence $\hat{x}_1, \hat{x}_2, \ldots$, $\hat{x}_i \in \hat{\calX}$, according to the following recursive equations:
\begin{eqnarray}
t_i &=& t_{i-1} + \Delta(z_i), \qquad t_0 = 0,\\
v_i &=& (u_{t_{i-1}+1}, u_{t_{i-1}+2}, \ldots, u_{t_i}),\\
\hat{x}_i &=& f(z_i,v_i),\\
z_{i+1} &=& g(z_i,v_i).
\end{eqnarray}

The guessing game is thus as follows: Alice has an individual sequence $\bx$ of length $n$. Bob is equipped with a finite--state guessing machine and unlimited number of uniform i.i.d.\ bits. The machine consumes bits, while passing through states in $\calZ$ and producing output symbols in $\hat{\calX}$. When the output is of length $n$ Bob submits it to Alice and asks ``is the sequence $\hat{\bx}$ within distortion $nD$ of $\bx$?" If the answer is affirmative, the game ends. Otherwise, the machine is restarted to its initial state (say, some $z_1 \in \calZ$) and the game repeats.

For a finite--state guessing machine F, denote by $G_F(\bx)$ the (random, due to the random input to the machine) number of guesses until $\bx$ is guessed with distortion at most $nD$. Our main result in this context is the following theorem, which, on the one hand, lower bounds the $\rho$--th moment of $G_F(\bx)$, yet, on the other, states that this lower bound is asymptotically achievable with a guessing distribution which can be implemented with a finite--state machine whose number of states is independent of the length $n$.

To present the result, we first remind the reader of the notion of finite--state compressibility \cite{ZL78}. We briefly follow the notation in \cite[Section III.A]{Merhav20a}. Let $L_E(b^n)$ be the total length of the binary strings $b_1,\ldots, b_n$ produced by an information lossless, finite--state encoder $E$ with at most $K$ states as it processes the string $\hat{\bx}$. The compression ration of $E$ is defined by $\rho_E(\hat{\bx}) = \frac{L_E(b^n)}{n}$, and the finite--state compresibility of $\hat{\bx}$ is defined by 
\begin{equation}
    \rho_K(\hat{\bx}) = \min_{E \in \calE(K)}\rho_E(\hat{\bx}), 
\end{equation}
where $\calE(K)$ is the set of all information lossless finite--state encoders with at most $K$ states.
\begin{theorem}\label{individual seq. theorem}
Denote by $\bE\{G_F^\rho|\bx\}$ the $\rho$--th moment of the guesswork for a finite--state machine F with at most $s$ states, given the individual sequence $\bx$. Then:

\emph{Converse part.} (i) For any fixed $s$ independent of $n$ we have
\begin{equation}\label{converse for individual (i)}
    \bE\{G_F^\rho|\bx\} \gexe \left[\sum_{\{\hat{\bx}:~d(\bx,\hat{\bx})\le nD\}} 2^{-LZ(\hat{\bx})}\right]^{-\rho}.
\end{equation}
(ii) Alternatively, we have
\begin{equation}\label{converse for individual (ii)}
    \bE\{G_F^\rho|\bx\} \gexe \left[  \sum_{\{\hat{\bx}:~d(\bx,\hat{\bx})\le nD\}} \exp_2\left\{-n \left(\rho_{K(l)}(\hat{\bx}) -\frac{\log\left(2 s^3 e\right)}{l}\right) \right\} \right]^{-\rho},
\end{equation}
for any $l$ which divides $n$ and $K(l)$ denoting the number of states of a machine implementing a Shannon code on blocks of size $l$. 

\emph{Direct part.} (i) The lower bound in \eqref{converse for individual (i)} is achieved using the guessing distribution \eqref{tP with LZ}. (ii) Assume $l$ divides $n$ and consider the distribution
\begin{equation}\label{block-LZ}
\tP(\hat{\bx}) = \prod_{i=0}^{n/l-1}\left[\frac{2^{-LZ({\hat{x}}_{il+1}^{il+l})}}{\sum_{\hat{x}^l \in \hat{\calX}^l}2^{-LZ(\hat{x}^l)}} \right],
\end{equation}
which is implementable with a finite--state machine with at most $l\alpha^l$ states for any $n$. Then 
\begin{equation}\label{direct for individual seq. blocks}
\bE\{G_{LZ}^\rho|\bx\} \lexe  \Bigg[\sum_{\{\hat{\bx}:~d(\bx,\hat{\bx})\le nD\}} \exp_2\bigg\{ - n\Big[ \rho_{K}(\hat{\bx}) + \frac{\log(4K^2)}{(1-\epsilon(l))\log l}+ \frac{K^2\log(4K^2)}{l} +\epsilon(l)\Big]  \bigg\}\Bigg]^{-\rho},
\end{equation}
where $\epsilon(l) \to 0$ as $l \to \infty$.
\end{theorem}
The complete proof of \Cref{individual seq. theorem} is given in \Cref{individual sequences}.

At this point, a few remarks are in order. First, note that the universal guessing distribution in \eqref{tP with LZ} is thus not only five-fold universal, it is also asymptotically optimal for \emph{four different setups:} guessing memoryless sources, guessing memoryless sources through a noisy channel, guessing sources with memory and guessing an individual sequence. Moreover, while this distribution is not implementable with a finite--state machine as $n$ grows without bound, it has both efficient algorithms which allow easily sampling from this distribution\footnote{This can be done without directly computing the complex sum in the denominator, but, rather, feeding an LZ decoder with uniform i.i.d.\ bits \cite[Section V.D]{MerhavCohen20}. Thus, these algorithms implement \eqref{tP with LZ} directly and efficiently, without the need for the block-wise implementation in \eqref{block-LZ}.}, as well as a block--wise version \eqref{block-LZ} which still asymptotically achieves the lower bound. 
\section{Three Lemmas on the Probability of $\calS(\bx)$ Under Universal Guessing Distributions}\label{sec. three lemmas}
When proving direct (achievability) results, the focus is on designing good \emph{guessing} distributions, and then analysing the expected guesswork under the relevant \emph{source} distribution. Thus, roughly speaking, a direct result includes three steps: introducing a guessing distribution $\tP(\hat{\bx})$, evaluating the probability of hitting the distortion-achieving set, $\tP[\calS(\bx)]$, and, finally, taking into account the actual source distribution $P$, computing the guesswork moment, $\bE\{G^\rho\} = \bE_P \bE\{G^\rho | \bx\}$. 

Evaluating $\tP[\calS(\bx)]$ is a key technical challenge. In this section, we include three lemmas which indeed evaluate (or at least lower bound) $\tP[\calS(\bx)]$ for the three major scenarios in this paper: memoryless sources (\Cref{First Lemma on PS}), noisy guessing (\Cref{Third Lemma on PS}) and sources with memory (\Cref{Second Lemma on PS}).  In fact, in the first and third scenarios, these lemmas will almost directly result in the achievability proofs. For sources with memory, the analysis of the achievability is more involved and will require additional steps. Interestingly, the three lemmas also have a similar structure, relating the exponential rate of the hitting probability to some rate--distortion--like function, depending on the scenario considered.
\subsection{Memoryless Sources}
For memoryless sources, we have the following.
\begin{lemma}\label{First Lemma on PS}
Assume sequences of length $n$ are chosen independently at random according to the distribution $\tP(\hat{\bx})\exe 2^{-n\hH_{\hat{\bx}}(\hat{X})}$ in \eqref{universal distribution for memoryless}. Then, for any $\bx \in \calX^n$
\begin{equation}
    \tP[\calS(\bx)] \exe 2^{-nR(D,\hat{P}_{\bx})},
\end{equation}
where $R(D,\hat{P}_{\bx})$ is the rate distortion function of $\hat{P}_{\bx}$.
\end{lemma}
\begin{proof}
\begin{eqnarray}
	\tP[\calS(\bx)]&=&\sum_{\{\hat{\bx}:~d(\bx,\hat{\bx})\le nD\}}\tP(\hat{\bx})\\
	&\exe&\sum_{\{\calT(Q_{\hat{X}|X}|\bx):~\bE_Qd(X,\hat{X})\le D\}}|\calT(Q_{\hat{X}|X}|\bx)|\cdot 2^{-n\hH_{\hat{\bx}}(\hat{X})}\\
	&\exe& \max_{\{Q_{\hat{X}|X}:~\bE_Qd(X,\hat{X})\le D\}} 2^{nH_Q(\hat{X}|X)}\cdot 2^{-nH_Q(\hat{X})}\label{type for specific x}\\
	&=& \max_{\{Q_{\hat{X}|X}:~\bE_Qd(X,\hat{X})\le D\}} 2^{-nI_Q(X;\hat{X})}\\
	&=& \exp_2\left\{-n \min_{\{Q_{\hat{X}|X}:~\bE_Qd(X,\hat{X})\le D\}} I_Q(X;\hat{X})\right\}\\
	&=&2^{-nR(D,\hat{P}_{\bx})},
\end{eqnarray}
where the last equality is since the minimum is taken over all conditional types $Q_{\hat{X}|X}$, conditioned on the specific $\bx$, satisfying the distortion constraint.
\end{proof}
\Cref{First Lemma on PS} is the key step in proving the direct result for memoryless sources. In fact, using \Cref{geometric lemma} of \Cref{sec. statement} together with \Cref{First Lemma on PS}, the direct result in \Cref{Direct for memoryless} easily follows:
\begin{proof}[Proof of \Cref{Direct for memoryless}]
\begin{eqnarray}
\bE\{G^\rho\}&=&\sum_{\bx}P(\bx)\cdot \bE\{G^\rho|\bx\} \label{first equation in proof of memoryless direct}\\
&\exe& \sum_{\bx} 2^{-n\left(\hH_{\bx}(X)+D(\hat{P}_{\bx} \| P)\right)} \cdot 2^{n\rho R(D,\hat{P}_{\bx})}\\
&=& \sum_{Q \in \calP_n} |\calT_Q| 2^{-n\left(H_Q(X)+D(Q \| P)\right)} \cdot 2^{n\rho R(D,Q)} \\
&\lexe& \sum_{Q \in \calP_n} 2^{nH_Q(X)} \cdot 2^{-n\left(H_Q(X)+D(Q \| P)\right)} \cdot 2^{n\rho R(D,Q)}\\
&\exe& \exp_2\left\{n\max_Q[\rho R(D,Q)-D(Q\|P)]\right\} \label{last equation in proof of memoryless direct}.
\end{eqnarray}
\end{proof}
\subsection{Noisy Guessing}\label{subsec lemma for noisy}
Analogously to \Cref{First Lemma on PS}, we now wish to evaluate the probability of $\calS(\by)$ under the guessing distribution we use. However, since guessing is done through a noisy channel, each guess induces a \emph{distribution} on the reconstruction words. A fortiori, when using a guessing distribution. The following lemma evaluates the probability of $\calS(\by)$ under the distribution induced by the guessing distribution \eqref{universal distribution for memoryless}.

\begin{lemma}\label{Third Lemma on PS}
Assume that sequences of length $n$ are chosen independently at random according to the distribution $\tP(\bx)\exe 2^{-n\hH_{\bx}(X)}$ in \eqref{universal distribution for memoryless}, then sent through the memoryless channel $W(\by|\bx)$. Denote by $Q$ the distribution induced on the $\by$--vectors. Then,
\begin{equation}
    Q[\calS(\by)] \exe \exp_2\{-n\hat{R}_W(D,\hat{P}_{\by})\},
\end{equation}
where $\hat{R}_W(D,\hat{P}_{\by})$ is given in \eqref{def. hatR}.
\end{lemma}
\begin{proof}
Guessing using the distribution in \eqref{universal distribution for memoryless} induces the following distribution on the $\by$--vectors:
\begin{eqnarray}
Q(\by)&\exe&\sum_{\bx\in\calX^n}\exp_2\{-n\hat{H}_{\bx}(X)\}\cdot
W(\by|\bx)\\
&\exe&\sum_{\calT(Q_{X|Y}|\by)}|\calT(Q_{X|Y}|\by)|\cdot
\exp_2\{-n[H_Q(X)+H_Q(Y|X)+D(Q_{Y|X}\|W|Q_X)]\}\\
&\exe&\max_{Q_{X|Y}}\exp_2\{n[H_Q(X|Y)-H_Q(X)-H_Q(Y|X)-D(Q_{Y|X}\|W|Q_X)]\}\\
&=&\exp_2\{-nH_Q(Y)\}\cdot\max_{Q_{X|Y}}\exp_2\{-nD(Q_{Y|X}\|W|Q_X)\}\\
&=&\exp_2\{-n[H_Q(Y)+\Gamma(Q_Y)]\}.
\end{eqnarray}
Note that $Q$ above is the distribution whose $Y$--marginal is $\hat{P}_{\by}$, hence $H_Q(Y)$ is, in fact, $\hat{H}_{\by}(Y)$. Thus, given that $\bY=\by$, the probability of a single successful guess within distortion $D$ is given by
\begin{align}
	\sum_{\{\by':~d(\by,\by')\le nD\}}&Q(\by') \\
	&\exe
	\sum_{\{\by':~d(\by,\by')\le nD\}}\exp_2\{-n[H_Q(Y')+\Gamma(Q_{Y'})]\}\\
	&\exe\max_{\{Q_{Y'|Y}:~\bE_Qd(Y,Y')\le D\}}|\calT(Q_{Y'|Y}|\bY=\by)|\cdot
	\exp_2\{-n[H_Q(Y')+\Gamma(Q_{Y'})]\}\\
	&\exe\max_{\{Q_{Y'|Y}:~\bE_Qd(Y,Y')\le D\}}
	\exp_2\{nH_Q(Y'|Y)\}\cdot\exp_2\{-n[H_Q(Y')+\Gamma(Q_{Y'})]\}\\
	&\exe\exp_2\left\{-n\min_{\{Q_{Y'|Y}:~\bE_Qd(Y,Y')\le D\}}
	[I_Q(Y;Y')+\Gamma(Q_{Y'})]\right\}\\
	&=\exp_2\{-n\hat{R}_W(D,Q_Y)\}.
\end{align}
Remembering that in the above $Q_Y$ is empirical distribution of the specific $\by$, that is, $\hat{P}_{\by}$, completes the proof.
\end{proof}

The direct result for noisy guessing under distortion now easily follows.
\begin{proof}[Proof of \Cref{theorem noisy} (Direct Part)]
Applying \Cref{geometric lemma} on the result in \Cref{Third Lemma on PS}, we have
\begin{equation}
\bE\{G^\rho|\bY=\by\}\exe\exp_2\{n\rho\hat{R}_W(D,\hat{P}_{\by})\}.
\end{equation}
The proof now follows the same steps as in \eqref{first equation in proof of memoryless direct}--\eqref{last equation in proof of memoryless direct}, resulting in 
\begin{align}
	\bE\{G^\rho\} &=\sum_{\by}P(\by)\cdot \bE\{G^\rho|\bY=\by\}
	\\
	&\exe
	\exp_2\{n\max_{Q_Y}[\rho\hat{R}_W(D,Q_Y)-D(Q_Y\|P)]\},
\end{align}
which completes the proof of this achievability result.
\end{proof}
\subsection{Sources with Memory}\label{subsec. lemma for memory}
Before presenting the main lemma in this section, a few definitions are required.

With proper scaling and gaps, $\Psi$-mixing sources can be approximated by block-memoryless sources, for which one can use the method of types over the super--alphabet of the blocks. Note, however, that directly applying the method of types to consecutive blocks of source symbols will not result in a good approximation, as the blocks are not independent, hence gaps between the blocks should be introduced. To make this precise, assume a source $n$-tuple is divided to consecutive, non-overlapping tuples of length $K+k$. We assume both $K$ and $k$ are fixed, yet $K \gg k$. We further assume that $K+k$ divides $n$ and denote $\frac{n}{K+k}=m$. Hence,
\begin{align}\label{x as blocks}
x_1^n &= x_1^K, x_{K+1}^{K+k}, \ldots, x_{n-K-k+1}^{n-k},x_{n-k+1}^{n}
\\
&\dfn  \ux_1, \ug_1, \ldots, \ux_{m},\ug_{m}, \quad \ux_i \in \calX^K, \ug_i \in \calX^k, 1 \leq i \leq m,
\end{align}
where we refer to the $K$-tuples $\ux_i$ as \emph{blocks}, and to the $k$-tuples $\ug_i$ as \emph{gaps}. Analogously to the type of a memoryless sequence, herein, we define the (\emph{generalized}, due to the gaps) \emph{$K$-type} of a sequence $x_1^n$ as the vector of empirical frequencies for each possible block of size $K$ in $x_1^n$, disregarding the content of the gaps. That is, denoting by $N_{\bx}(a_1^K)$ the number of occurrences of  $a_1^K \in \calX^K$ in the blocks $\{\ux_i\}_{i=1}^{m}$ of $\bx$, we have
\begin{equation}
\hat{P}^K_{\bx}  \dfn \left( \frac{N_{\bx}(a_1^K)}{m}, a_1^K \in \calX^K \right).
\end{equation}
The $K$-type class of  $P^K$, denoted by $T^K(P^K)$, is the set of all sequences with the same $K$-type $P^K$, that is, $T^K(P^K) = \left\{\bx \in \calX^n : \hat{P}^K_{\bx} = P^K\right\}$. Finally, denoting by $\calT^K$ the set of all possible $K$-types $P^K$, clearly, 
\begin{equation}\label{number of types}
\left|\calT^K\right| \leq \left( m +1\right)^{|\calX|^K}.
\end{equation}

The following lemma generalizes \Cref{First Lemma on PS} to guessing sequences with memory based on LZ parsing, via the distribution in \eqref{tP with LZ}.
\begin{lemma}\label{Second Lemma on PS}
Assume sequences of length $n$ are chosen independently at random according to the distribution $\tP(\hat{\bx})\exe 2^{-LZ(\hat{\bx})}$ in \eqref{tP with LZ}. Then, for any $\delta,\epsilon > 0$ there exists a sequence $\epsilon_{m} \to 0$ as $m \to \infty$ such that 
\begin{equation}
        \tP[\calS(\bx)] \gexe 2^{-\nu_{K+k}(n)}(1-2\epsilon_{m})\exp_2\left\{-m\left[R_K(D',\hat{P}^K_{\bx})+\epsilon\right]\right\},
\end{equation}
where $D' = D-\delta_2 -\frac{\delta}{K}$, $\delta_2 = \frac{k(D_{max}-D)}{K}$, $\hat{P}^K_{\bx}$ is the $K$-type of $\bx$, and for any fixed $K$ and $k<K$, $\lim_{n \to \infty} \frac{1}{n}\nu_{K+k}(n) = O\left(\frac{1}{K}\right)$.
\end{lemma}
To prove \Cref{Second Lemma on PS} we first, in \Cref{prop. exponential bound for block memoryless} below, consider a specific \emph{block--memoryless distribution}, induced by the distribution which achieves the rate--distortion function $R_K(D,\hat{P}^K_{\bx})$, and connect the probability of hitting the distortion--achieving blocks under this distribution to the rate--distortion function. In a sense, this relates the probability of a slightly smaller set, $\underbar{\calS}(\bx) \subset \calS(\bx)$, to be defined precisely later, to the rate-distortion function of a block memoryless source. Then, we upper bound this block--memoryless distribution using the code length of the LZ encoder. Finally, we connect the two results to prove \Cref{Second Lemma on PS}.

\begin{proposition}\label{prop. exponential bound for block memoryless}
Fix $\bx \in \calX^n$ with a $K$-type $\hat{P}^K_{\bx}$. For any $\delta >0$, set $D' = D-\delta_2-\frac{\delta}{K}$, with $\delta_2 = \frac{k(D_{max}-D)}{K}$ and denote by $P^K(\hat{x}|x)$, with $\hat{x} \in \hat{\calX}^K$ and $x \in \calX^k$ the conditional distribution achieving $R_K(D',\hat{P}^K_{\bx})$. Finally, denote by $Q^K$ the resulting distribution on the construction alphabet $\hat{\calX}^K$, that is, $Q^K(\hat{x}) = \sum_{x}P^K(\hat{x}|x)\hat{P}^K_{\bx}(x)$. For any $\epsilon>0$, the following holds  
\begin{equation}
\sum_{\hat{\bs}_1^{m} \in (\hat{\calX}^K)^{m}:~\sum_i{d(\ux_i,\hat{\bs}_i)} \le mK(D-\delta_2)} \prod_{i=1}^{m}Q^K\left(\hat{\bs}_i\right) \ge (1-2\epsilon_{m})\exp_2\left\{-m\left[R_K(D',\hat{P}^K_{\bx})+\epsilon\right]\right\}.
\end{equation}
\end{proposition}
\begin{proof}
We first state a result for memoryless sources, then extend it to a block--memoryless scenario, minding the gaps between the blocks. 

Fix $\bx \in \calX^n$ with a type $\hat{P}_{\bx}$. For any $\delta >0$, set $\Dm = D-\delta$ and denote by $P(\hat{x}|x)$ the conditional distribution achieving $R(\Dm,\hat{P}_{\bx})$. Finally, denote by $Q$ the resulting distribution on the construction alphabet, that is, $Q(\hat{x}) = \sum_{x}P(\hat{x}|x)\hat{P}_{\bx}(x)$. For any $\epsilon>0$, the following holds  
\begin{equation}\label{eq. prop on Q - memoryless}
\sum_{\hat{\bx} \in \calS(\bx)}Q^n(\hat{\bx}) \ge (1-2\epsilon_n)\exp_2\left\{-n\left[R(\Dm,\hat{P}_{\bx})+\epsilon\right]\right\},
\end{equation}
where $Q^n(\hat{\bx}) = \prod_{i=1}^n Q(\hat{x})$ and $\epsilon_n \to 0$ as $n \to \infty$ for all $\bx \in \calX^n$. To prove \eqref{eq. prop on Q - memoryless}, we follow \cite[eq. (3.2.8)-(3.2.10)]{berger1971rate}. Define
\begin{equation}
\calT(\bx) = \left\{ \hat{\bx} : \log\frac{P^n(\hat{\bx} | \bx)}{Q^n(\hat{\bx})} \leq n \left[R(\Dm,\hat{P}_{\bx})+\epsilon\right]\right\},
\end{equation}
where $P^n(\hat{\bx}|\bx) = \prod_{i=1}^{n}P(\hat{x}_i|x_i)$, $P(\hat{x}_i|x_i)$ and $Q(\hat{x})$ being the optimal test channel reconstruction distribution for $R(\Dm,\hat{P}_{\bx})$, respectively. We have
\begin{align}
\sum_{\hat{\bx} \in \calS(\bx)}Q^n(\hat{\bx}) &\ge \sum_{\hat{\bx} \in \calS(\bx) \cap \calT(\bx)}Q^n(\hat{\bx})\\
&\ge \exp_2\left\{-n\left[R(\Dm,\hat{P}_{\bx})+\epsilon\right]\right\}\sum_{\hat{\bx} \in \calS(\bx) \cap \calT(\bx)}P^n(\hat{\bx}|\bx)\\
&\ge \exp_2\left\{-n\left[R(\Dm,\hat{P}_{\bx})+\epsilon\right]\right\} \left[ \sum_{\hat{\bx} \in \calS(\bx)}P^n(\hat{\bx}|\bx) - \sum_{\hat{\bx} \in \calT^c(\bx)}P^n(\hat{\bx}|\bx) \right].
\end{align}
To conclude the proof of \eqref{eq. prop on Q - memoryless}, we have to show that there exists a sequence $\epsilon_n \to 0$ such that both $\sum_{\hat{\bx} \in \calS(\bx)}P^n(\hat{\bx}|\bx) \ge 1-\epsilon_n$ and $\sum_{\hat{\bx} \in \calT^c(\bx)}P^n(\hat{\bx}|\bx) \leq \epsilon_n$ uniformly in $\bx$. To this end, we have the following two claims, whose proofs are given in \Cref{app proofs}.
\begin{claim}\label{claim on sum over S}
There exists $\epsilon'_n$ such that for all $\bx$, $\sum_{\hat{\bx} \in \calS(\bx)}P^n(\hat{\bx}|\bx) \ge 1-\epsilon'_n$ with $\epsilon'_n \to 0$ as $n \to \infty$.
\end{claim}
\begin{claim}\label{claim on sum over tc}
There exists $\epsilon''_n$ such that for all $\bx$, $\sum_{\hat{\bx} \in \calT^c(\bx)}P^n(\hat{\bx}|\bx) \leq \epsilon''_n$ with $\epsilon''_n \to 0$ as $n \to \infty$.
\end{claim}
Taking $\epsilon_n = \max(\epsilon'_n,\epsilon''_n)$ completes the proof of \eqref{eq. prop on Q - memoryless}.

To complete the proof of \Cref{prop. exponential bound for block memoryless}, note that \eqref{eq. prop on Q - memoryless} can be easily extended to \emph{block--memoryless} sources, simply by applying it to sequences of length $m$ over $\calX^K$, with the proper definition on a type over  $\calX^K$ and output distribution $Q^K$ on $\hat{\calX}^K$. However, we take this a step further, and apply it to a sequence $\bx$ of length $n$ over $\calX$, using a \emph{memoryless distribution} over $K$-tuples and the $K$-type of $\bx$. Namely, given any $\bx \in \calX^n$, apply \eqref{eq. prop on Q - memoryless} on the sequence $(\ux_1,\ldots,\ux_m)$ (that is, only the $K$-tuple blocks of $\bx$, without the $k$-tuple gaps). This sequence, which includes $m$ blocks of length $K$ over $\calX$, when viewed as an $m$-tuple over $\calX^K$ has a type $\hat{P}^K_{\bx}$. Thus, setting a per-super-letter distortion level $K(D-\delta_2)$, the proposition follows. 
\end{proof}

Using \Cref{prop. exponential bound for block memoryless} above, we can now prove \Cref{Second Lemma on PS}.
\begin{proof}{( \Cref{Second Lemma on PS})}
Let $M^{K+k}(\cdot)$ be any probability distribution on $\hat{\calX}^{K+k}$ and \begin{equation}
\prod_{i=1}^{m}M^{K+k}\left(\hat{x}_{(i-1)(K+k)+1}^{i(K+k)}\right)
\end{equation}
be the product distribution on $\hat{\calX}^n$. We first have \Cref{claim ZIv's inequality} below, which can be viewed as an extension of Ziv's inequality \cite[Theorem 1]{ZL78}, and is proved in \Cref{app proofs}.
\begin{claim}\label{claim ZIv's inequality}
For any sequence $\hat{\bx} \in \hat{\calX}^n$ and any distribution $M^{K+k}(\cdot)$ on $\hat{\calX}^{K+k}$, the LZ code length (in bits), $LZ(\hat{\bx})$, satisfies
\begin{equation}
LZ(\hat{\bx}) \leq -\log \left[\prod_{i=1}^{m}M^{K+k}\left({\hat{x}}_{(i-1)(K+k)+1}^{i(K+k)}\right)\right] + \nu_{K+k}(n),
\end{equation}
where $\nu_{K+k}(n)$ depends on $|\hat{\calX}|$ yet satisfies $\lim_{n \to \infty} \frac{1}{n}\nu_{K+k}(n) = O\left(\frac{1}{K}\right)$ for any fixed $K$ and $k<K$.
\end{claim}

Similar to \eqref{x as blocks}, consider the partition of a sequence $\hat{\bx}$ of length $n$ over $\hat{\calX}$ into non-overlapping blocks of length $K$, followed by gaps of length $k$
\begin{equation}
\hat{\bx} = \uhx_1, \uhg_1, \ldots, \uhx_{m},\uhg_{m}.
\end{equation}
Next, define $\underline{\calS}(\bx) \subset \calS(\bx)$ as follows
\begin{equation}\label{def. underline S}
\underline{\calS}(\bx)  = \left\{\hat{\bx}: \uhg_i = (\hat{x}_0, \ldots, \hat{x}_0) \text{ for all } i,~\sum_i{d(\ux_i,\uhx_i)}\le mK(D-\delta_2)\right\}.
\end{equation}
That is, $\underbar{\calS}(\bx)$ includes only sequences in $\calS(\bx)$ for which the gaps include only $\hat{x}_0$ (for some fixed symbol $\hat{x}_0$), yet the distortion constraint is satisfied. Consequently, we have 
\begin{align}
\tP[\calS(\bx)] &\ge \sum_{\hat{\bx} \in \calS(\bx)} 2^{-LZ(\hat{\bx})}\\
&\ge 2^{-\nu_{K+k}(n)}\sum_{\hat{\bx} \in \calS(\bx)} \prod_{i=1}^{m}M^{K+k}\left(\hat{x}_{(i-1)(K+k)+1}^{i(K+k)}\right)\label{test}\\
&\ge 2^{-\nu_{K+k}(n)}\sum_{\hat{\bx} \in \calS(\bx)} \prod_{i=1}^{m}Q^K\left(\hat{x}_{(i-1)(K+k)+1}^{i(K+k)-k}\right)\mathbbm{1}_{\left\{ \hat{x}_{i(K+k)-k+1}^{i(K+k)} = (\hat{x}_0, \ldots \hat{x}_0)\right\}}\\
&= 2^{-\nu_{K+k}(n)}\sum_{\hat{\bx} \in \underline{\calS}(\bx)} \prod_{i=1}^{m}Q^K\left(\hat{x}_{(i-1)(K+k)+1}^{i(K+k)-k}\right)\\
&= 2^{-\nu_{K+k}(n)}\sum_{\hat{\bs}_1^{m} \in (\hat{\calX}^K)^{m}:~\sum_i{d(\ux_i,\hat{\bs}_i)} \le mK(D-\delta_2)} \prod_{i=1}^{m}Q^K\left(\hat{\bs}_i\right)\\
&\ge 2^{-\nu_{K+k}(n)}(1-2\epsilon_{m})\exp_2\left\{-m\left[R_K(D',\hat{P}^K_{\bx})+\epsilon\right]\right\},
\end{align}
where \eqref{test} is by \Cref{claim ZIv's inequality} and the last inequality follows from \Cref{prop. exponential bound for block memoryless}.
\end{proof}
\section{Converse for Noisy Guesses}\label{sec. converse for noisy}
The direct part of \Cref{theorem noisy} was proved in \Cref{subsec lemma for noisy} via \Cref{Third Lemma on PS}. In this section, we prove the converse part of \Cref{theorem noisy}. We begin from a simple preparatory lemma.
\begin{lemma}\label{lemma for W(S(y)|x)}
Let $\calT(Q_Y)$ be a given type
class of $\by$--vectors, and let $\bx\in\calT(Q_X)$ be given. Then,
\begin{equation}
	\frac{1}{|\calT(Q_Y)|}\sum_{\by\in\calT(Q_Y)}W[\calS(\by)|\bx]\lexe
	\exp_2\{-n\hat{R}_W(D,Q_Y)\}~~~~~~\forall~\bx\in\calX^n.
\end{equation}
\end{lemma}
Note that the lemma above is, in a sense, a matching upper bound to the result in \Cref{Third Lemma on PS}. However, while \Cref{Third Lemma on PS} assumes that the vector $\bx$ is drawn from the universal distribution used in the direct results, \Cref{lemma for W(S(y)|x)}, which is later utilized in the \emph{converse result}, does not assume any distribution on $\bx$. Yet, it includes averaging over $\by\in\calT(Q_Y)$.
\begin{proof}
\begin{align}
	\frac{1}{|\calT(Q_Y)|}&\sum_{\by\in\calT(Q_Y)}W[\calS(\by)|\bx]\\
	&=\frac{1}{|\calT(Q_Y)|}\sum_{\by\in\calT(Q_Y)}
	\sum_{\{\by':~d(\by,\by')\le nD\}} W(\by'|\bx)\\
	&=\frac{1}{|\calT(Q_Y)|}\sum_{\{\calT(Q_{YY'|X}|\bx):~\bE_Qd(Y,Y')\le D\}}|\calT(Q_{YY'|X}|\bx)|\\
	& \hspace{5cm} \cdot
	\exp_2\{-n[H_Q(Y'|X)+D(Q_{Y'|X}\|W|Q_X)]\}\\
	&\exe\exp_2\Bigg\{-n\min_{\{Q_{YY'|X}:~\bE_Qd(Y,Y')\le D\}}[H_Q(Y)-H_Q(Y,Y'|X)\\
	& \hspace{5cm} +
	H_Q(Y'|X)+D(Q_{Y'|X}\|W|Q_X)]\Bigg\}\\
	&=\exp_2\left\{-n\min_{\{Q_{YY'|X}:~\bE_Qd(Y,Y')\le D\}}[H_Q(Y)-H_Q(Y|X,Y')+
	D(Q_{Y'|X}\|W|Q_X)]\right\}\\
	&=\exp_2\left\{-n\min_{\{Q_{YY'|X}:~\bE_Qd(Y,Y')\le D\}}[I_Q(Y;X,Y')+
	D(Q_{Y'|X}\|W|Q_X)]\right\}\\
	&\le\exp_2\Bigg\{-n\min_{\{Q_{Y'|Y}:~\bE_Qd(Y,Y')\le D\}}[I_Q(Y;Y')\\
	& \hspace{5cm} +
	\min_{\{Q_{Y'|X}:(Q_X\odot Q_{Y'|X})_{Y'}=Q_{Y'}\}}
	D(Q_{Y'|X}\|W|Q_X)]\Bigg\}\\
	&=\exp_2\left\{-n\min_{\{Q_{Y'|Y}:~\bE_Qd(Y,Y')\le D\}}[I_Q(Y;Y')+\Gamma(Q_X,Q_{Y'})]\right\}\\
	&\le\exp_2\left\{-n\min_{\{Q_{Y'|Y}:~\bE_Qd(Y,Y')\le D\}}[I_Q(Y;Y')+\Gamma(Q_{Y'})]\right\}\\
	&=\exp_2\{-n\hat{R}_W(D,Q_Y)\}.
\end{align}
\end{proof}

The proof of the converse is now as follows.
\begin{proof}[Proof of \Cref{theorem noisy} (Converse Part)]
Denote by $k(Q_Y)$ a positive integer, whose value, possibly depending on the type of $\by$, $Q_Y$, will be defined later. We have the following chain of inequalities, whose explanations are given below. 
\begin{eqnarray}
\bE\{G^\rho\}&=&\sum_{Q_Y}P[\calT(Q_Y)]\bE\{G^\rho|\by\in\calT(Q_Y)\}\\
& \exe &\sum_{Q_Y}2^{-nD(Q_Y || P)}\bE\{G^\rho|\by\in\calT(Q_Y)\}\\
& \ge &\sum_{Q_Y}2^{-nD(Q_Y || P)}k(Q_Y)^\rho \cdot\mbox{Pr}\{G>
k(Q_Y)|\by\in\calT(Q_Y)\}\label{converse-noisy-a} \\
& \ge &\sum_{Q_Y}2^{-nD(Q_Y || P)}k(Q_Y)^\rho \sum_{\by\in\calT(Q_Y)} \frac{1}{|\calT(Q_Y)|}\mbox{Pr}\{G>
k(Q_Y)|\by\}\label{converse-noisy-b} \\
& \ge &\sum_{Q_Y}2^{-nD(Q_Y || P)}k(Q_Y)^\rho \sum_{\by\in\calT(Q_Y)} \frac{1}{|\calT(Q_Y)|}\left(1-\sum_{i=1}^{k(Q_Y)} W(\calS(\by)|\bx_i)\right)\label{converse-noisy-c} \\
& = &\sum_{Q_Y}2^{-nD(Q_Y || P)}k(Q_Y)^\rho \left[ 1- \sum_{i=1}^{k(Q_Y)}\sum_{\by\in\calT(Q_Y)}\frac{1}{|\calT(Q_Y)|} W(\calS(\by)|\bx_i) \right] \\
& \gexe &\sum_{Q_Y}2^{-nD(Q_Y || P)}k(Q_Y)^\rho \left[ 1- \sum_{i=1}^{k(Q_Y)} 2^{-n\hat{R}_W(D,Q_Y)}  \right]\label{converse-noisy-d} \\
& = &\sum_{Q_Y}2^{-nD(Q_Y || P)}k(Q_Y)^\rho \left[ 1- k(Q_Y) 2^{-n\hat{R}_W(D,Q_Y)}  \right] \\
& \exe &\sum_{Q_Y}2^{-nD(Q_Y || P)}2^{\rho n\left(\hat{R}_W(D,Q_Y)-\epsilon\right)} \left[ 1- 2^{-n\epsilon}  \right]\label{converse-noisy-e} \\
&  \exe & \exp_2\left\{ n \max_{Q_Y} \left[ \rho \hat{R}_W(D,Q_Y)-D(Q_Y || P) - \rho \epsilon\right]\right\}. 
\end{eqnarray}
In the above chain, \eqref{converse-noisy-a} is since $\bE G^\rho = \sum_{k=1}^{\infty} k^\rho \cdot\mbox{Pr}\{G=k\} \ge k^\rho \cdot\mbox{Pr}\{G >k\}$. \eqref{converse-noisy-b} is since $\bY$ is drawn from a DMS, hence the conditioning $\bY\in\calT(Q_Y)$ means that $\bY$ is uniformly
distributed within $\calT(Q_Y)$, which explains the uniform averaging across $\calT(Q_Y)$. \eqref{converse-noisy-c} is since $\mbox{Pr}\{G >k|\by\} = 1-\mbox{Pr}\{\cup_{i=1}^{k} ~ \text{guess i is successful}|\by\} \ge 1 - \sum_{i=1}^{k}\mbox{Pr}\{\text{guess i is successful}|\by\}$, and $\mbox{Pr}\{\text{guess i is successful}|\by\} = W(\calS(\by)|\bx_i)$. \eqref{converse-noisy-d} is due to \Cref{lemma for W(S(y)|x)}. \eqref{converse-noisy-e} is by choosing $k(Q_Y) = \ceil{ 2^{n\left(\hat{R}(D,Q_Y)-\epsilon\right)}}$ for some arbitrarily small $\epsilon$. Since the above is true for any $\epsilon> 0$, we have
\begin{equation}
\bE\{G^\rho\} \gexe \exp_2 \left\{ n \max_{Q_Y} \left[ \rho \hat{R}_W(D,Q_Y)-D(Q_Y || P) \right]\right\},
\end{equation}
which completes the proof of the converse.
\end{proof}

\section{Direct and Converse for Sources with Memory}\label{sec. memory}
\subsection{Direct}\label{parsing section}
We can now assert the following theorem, which gives an upper bound on the guesswork exponent, using a $K$-tuples parsing strategy.
\begin{theorem}\label{direct with K-tuples}
For any stationary $\Psi$-mixing source $P$ over $\calX$, if guesses are selected independently at random according to the distribution \eqref{tP with LZ}, then, for  $K \gg k$, the $\rho$-th guesswork moment, subject to a distortion constraint $D$, satisfies 
\begin{equation}
\limsup_{n\to\infty}\frac{1}{n}\log\bE\{G^\rho\} \lexe \liminf_{K\to\infty}\sup_{Q^K} \frac{1}{K} \left[\rho R_K(D,Q^K) - D\left(Q^K||P^K\right)\right].
\end{equation}
\end{theorem}
\begin{proof}
We first bound the expected guesswork moment in terms of the guessing distribution $\tP$ via \Cref{geometric lemma}, and further bound $\tP[\calS(\bx)]$ via \Cref{Second Lemma on PS}. Specifically,
\begin{align}
\bE\{G^\rho\}&=\bE\{\bE\{G^\rho|\bX\} \}\label{direct1}\\
&\exe \bE\left\{\tP[\calS(\bX)]^{-\rho}\right\}\label{direct2} \\
&=\sum_{\bx \in \calX^n}P(\bx)\tP[\calS(\bx)]^{-\rho}\label{direct3}\\
&\leq 2^{\rho\nu_{K+k}(n)} \sum_{\bx \in \calX^n}P(\bx)\frac{1}{(1-2\epsilon_{m})^\rho}\exp_2\left\{\rho m\left[R_K(D',\hat{P}^K_{\bx})+\epsilon\right]\right\}\label{direct5}\\
&= \frac{2^{\rho\nu_{K+k}(n)}}{(1-2\epsilon_{m})^\rho} \sum_{\bx \in \calX^n}P(\bx)\exp_2\left\{\rho m\left[R_K(D',\hat{P}^K_{\bx})+\epsilon\right]\right\},\label{direct55}
\end{align}
where \eqref{direct2} is due to \Cref{geometric lemma} and \eqref{direct5} is due to \Cref{Second Lemma on PS}. 

While it is tempting to replace the sum over $\bx \in \calX^n$ with the usual sum over types, unlike the memoryless case, sequences in $T^K(P^K)$ are not equiprobable. Fortunately, dividing $T^K(P^K)$ into subsets of sequences, which differ only in the content of the \emph{gaps between the blocks}, and with the right choice of $k$, results in asymptotically equiprobable subsets, whose probability can be described similarly to that of memoryless sequences. Specifically, for any $\bx \in \calX^n$, whose blocks are denoted by $\ux_i, 1 \leq i \leq m$ (see \eqref{x as blocks}), define the \emph{gaps oblivious} set as
\begin{equation}
G_{\bx}^K = \left\{ \tilde{\bx} \in \calX^n : \tilde{\ux}_i = \ux_i , 1 \leq i \leq m \right\}.
\end{equation}
That is, $G_{\bx}^K$ includes all sequences which differ from $\bx$ only in the gaps between the blocks. $G_{\bx}^K$ satisfies the following bounds, proved in \Cref{appendix proof of probability of gap oblovious}.
\begin{lemma}\label{probability of gap oblovious}
For any stationary $\Psi$-mixing source $Q$ and any $\bx \in \calX^n$,
\begin{align}
Q\left(G_{\bx}^K\right) &\leq (1+\epsilon_k)^{m-1} \exp_2\left\{-m \left(D\left(\hat{P}^K_{\bx}||Q^K\right) + \hat{H}_{\bx}^K(X^K)\right)\right\}
\\
Q\left(G_{\bx}^K\right) &\ge (1-\epsilon_k)^{m-1} \exp_2\left\{-m \left(D\left(\hat{P}^K_{\bx}||Q^K\right) + \hat{H}_{\bx}^K(X^K)\right)\right\},
\end{align}
where $Q^K()$ is the $K$-th order marginal distribution of $Q$, $\hat{H}_{\bx}^K(X^K)$ is the empirical entropy associated with the empirical distribution $\hat{P}^K_{\bx}$ and $\epsilon_k \to 0$ as $k \to \infty$.
\end{lemma}
Note that if the $K$-th order marginal of $Q$ satisfies $Q^K = \hat{P}^K_{\bx}$, the divergences in \Cref{probability of gap oblovious} are zero and we have,\footnote{$\hat{P}^K_{\bx}\left(G_{\bx}^K\right)$ is the probability of a \emph{set of sequences}, rather than the probability of a single sequence. Sequences in the subset are not necessarily equiprobable, yet the probability of the whole set is equal to that of \emph{other sets of sequences}, as long as the sets are defined for the same $K$-type $P^K$.}
\begin{align}
\hat{P}^K_{\bx}\left(G_{\bx}^K\right) &\leq (1+\epsilon_k)^{m-1} 2^{-m\hat{H}_{\bx}^K(X^K)}\label{P of G same type upper}
\\
\hat{P}^K_{\bx}\left(G_{\bx}^K\right) &\ge (1-\epsilon_k)^{m-1} 2^{-m\hat{H}_{\bx}^K(X^K)}.
\end{align}
Denote by $\left| T^K(P^K) \right|_G$ the number of \emph{distinct} sets $G_{\bx}^K$ in $T^K(P^K)$. The following corollary, proved in \Cref{appendix proof of size of generalized type}, bounds the number of such sets within a $K$-type class $T^K(P^K)$.
\begin{corollary}\label{size of generalized type}
$\left| T^K(P^K) \right|_G \leq (1-\epsilon_k)^{-m+1} 2^{m\hat{H}_{\bx}^K(X^K)}$.
\end{corollary}

We are now able to further bound $\bE\{G^\rho\}$. Continuing from \eqref{direct55},
\begin{align}
\bE\{G^\rho\}& \lexe \frac{2^{\rho\nu_{K+k}(n)}}{(1-2\epsilon_{m})^\rho}  \sum_{T^K \in \calT^K} \sum_{G_{\bx}^K \in T^K} P(G_{\bx}^K) 2^{\rho m\left[R_K(D',\hat{P}^K_{\bx})+\epsilon\right]}\label{direct6}\\
&\leq \frac{2^{\rho\nu_{K+k}(n)}(1+\epsilon_k)^{m-1}}{(1-2\epsilon_{m})^\rho}  \sum_{T^K \in \calT^K} \sum_{G_{\bx}^K \in T^K}  2^{m \left(-D\left(\hat{P}^K_{\bx}||P^K\right) - \hat{H}_{\bx}^K(X^K)\right)} 2^{\rho m\left[R_K(D',\hat{P}^K_{\bx})+\epsilon\right]}\label{direct7}\\
&\leq \frac{2^{\rho\nu_{K+k}(n)}(1+\epsilon_k)^{m-1}}{(1-\epsilon_k)^{m-1}(1-2\epsilon_{m})^\rho}  \\
& \hspace{3cm}\sum_{T^K \in \calT^K}   2^{m\hat{H}_{\bx}^K(X^K)}  2^{m \left(-D\left(\hat{P}^K_{\bx}||P^K\right) - \hat{H}_{\bx}^K(X^K)\right)} 2^{\rho m\left[R_K(D',\hat{P}^K_{\bx})+\epsilon\right]}\label{direct8}\\
&= \frac{2^{\rho\nu_{K+k}(n)}(1+\epsilon_k)^{m-1}}{(1-\epsilon_k)^{m-1}(1-2\epsilon_{m})^\rho}  \sum_{T^K \in \calT^K} 2^{-m D\left(\hat{P}^K_{\bx}||P^K\right)} 2^{\rho m\left[R_K(D',\hat{P}^K_{\bx})+\epsilon\right]}\label{direct9}\\
&\leq \frac{2^{\rho\nu_{K+k}(n)}(1+\epsilon_k)^{m-1}}{(1-\epsilon_k)^{m-1}(1-2\epsilon_{m})^\rho}  \left|\calT^K\right| \max_{\hat{P}^K_{\bx}} 2^{-m D\left(\hat{P}^K_{\bx}||P^K\right)} 2^{\rho m\left[R_K(D',\hat{P}^K_{\bx})+\epsilon\right]}\label{direct10}\\
&\leq \frac{2^{\rho\nu_{K+k}(n)}(1+\epsilon_k)^{m}\left( m +1\right)^{|\calX|^K}}{(1-\epsilon_k)^{m}(1-2\epsilon_{m})^\rho}   2^{\max_{\hat{P}^K_{\bx}} m\left[\rho R_K(D',\hat{P}^K_{\bx})+ \rho \epsilon - D\left(\hat{P}^K_{\bx}||P^K\right) \right]}\label{direct11}\\
&=  \exp_2 \Bigg\{ \max_{\hat{P}^K_{\bx}} n\bigg[\rho \frac{1}{K+k}R_K(D',\hat{P}^K_{\bx}) - \frac{1}{K+k}D\left(\hat{P}^K_{\bx}||P^K\right) \\
 & \hspace{0.8cm} + \rho\left(\epsilon-\frac{\log(1-2\epsilon_{m})-\nu_{K+k}(n)}{n}\right) + \frac{\log\left(\frac{1+\epsilon_k}{1-\epsilon_k}\right)}{K+k} + \frac{|\calX|^K \log\left(m+1\right)}{n}\bigg] \Bigg\} \label{direct12}.
\end{align}
In the above chain, \eqref{direct7} is due to \Cref{probability of gap oblovious}, while \eqref{direct8} is due to \eqref{P of G same type upper}; \eqref{direct11} follows from \eqref{number of types} and since $\frac{1-\epsilon_k}{1+\epsilon_k} \leq 1$. 

Now, fix $K \gg k$. Since $D' = D-\delta_2 -\frac{\delta}{K}$, with $\delta_2 = \frac{k(D_{max}-D)}{K}$, and since the above is true for any $\delta,\epsilon>0$, taking the limit over $n$,  we have
\begin{equation}\label{direct before moving to P^K}
\limsup_{n\to\infty}\frac{1}{n}\log\bE\{G^\rho\} \lexe \max_{\hat{P}^K_{\bx}} \frac{1}{K}\left[\rho R_K(D,\hat{P}^K_{\bx}) - D\left(\hat{P}^K_{\bx}||P^K\right) + O(1)\right],
\end{equation}
where we have also used the continuity of $R_K(D,\hat{P}^K_{\bx})$ in $D$. Finally, the r.h.s of \eqref{direct before moving to P^K} is at least as large if the maximization is done over all distributions on $\calX^K$, rather than only (fractional) types. Thus, enlarging the optimization domain, then taking a limit over $K$, we have
\begin{equation}
\limsup_{n\to\infty}\frac{1}{n}\log\bE\{G^\rho\} \lexe \liminf_{K\to\infty}\sup_{Q^K}\frac{1}{K} \left[\rho R_K(D,Q^K) - D\left(Q^K||P^K\right)\right],
\end{equation}
which completes the proof.
\end{proof}
\subsection{Converse}\label{section converse}
For a converse theorem, we do not assume that the guesser is restricted to any randomized guessing strategy. In fact, a guesser is allowed to use a \emph{guessing list}, $\calG_n$, which is an ordered list of all members of $\hat{\calX}^n$. That is, $\calG_n=\{\hat{\bx}_1,\hat{\bx}_2,\ldots,\hat{\bx}_{|\hat{\calX}|^n}\}$. $\calG_n$ is associated with a \emph{guessing function}, $G(\bx)$, which is the function that maps $\calX^n$ to the integers, by assigning each $\bx\in\calX^n$ the smallest integer $k$ for
which $d(\bx,\hat{\bx}_k) \leq nD$, $\hat{\bx}_k \in \calG_n$, namely, $\hat{\bx}_k$ is the first element of $\calG_n$ which is in $\calS(\bx)$. Thus, $G(\bx)$ is the number of guesses required until successfully meeting the distortion criteria, using $\calG_n$, when $\bX=\bx$. Note that any randomized guessing strategy can be implemented with such a list. Note also that for finite alphabet, and assuming that for all $x\in\calX$ we have $d_{\min}=\min_{\hat{x}\in \hat{\calX}}d(x,\hat{x})=0$, one can easily append any list $\calG_n$ such that the range of $G(\bx)$ is finite. Thus, the following converse bounds from below the best achievable exponent of the $\rho$-th guessing moment.
\begin{theorem}\label{theo. converse}
For any source $P$ with an $K$-letter marginal $P^K$ the guesswork $\rho$-th moment exponent at distortion level $D$ is lower bounded by
\begin{equation}
\liminf_{K \to\infty}\inf_{\calG_K}\frac{1}{K}\log \bE_{P^K}\{G^\rho(\bX)\} \ge \liminf_{K \to\infty} \sup_{Q^K}\frac{1}{K}\left[\rho R_K(D,Q^K) -D(Q^K\|P^K)\right],
\end{equation}
where $P^K$ is any distribution on $\calX^K$.
\end{theorem}
\begin{proof}
Following \cite{AM98}, the key tool in the converse is to view $\ceil*{-\log G(\bx)}$ as the length of a rate--distortion code for $\bx \in \calX^K$. Specifically, assume a guesser indeed has a list $\calG_K$ such that for any $\bx \in \calX^K$ there is at least one item in the list, say $\hat{\bx}_k$ such that $d(\bx,\hat{\bx}_k) \leq KD$. Then, this list can be viewed as a rate-distortion codebook. To encode a word $\bx$, the encoder will simply use the smallest integer $k$ for which $d(\bx,\hat{\bx}_k) \leq KD$, with an appropriate code for the integers. A decoder will simply output $\hat{\bx}_k$. Consider the following distribution on the positive integers:
\begin{equation}\label{dist_for_ints}
p_i = \frac{c(\delta)}{i^{1+\delta}}
\end{equation}   
where $\delta>0$ and 
\begin{equation}
c(\delta)= \left[\sum_{i=1}^{\infty} \frac{1}{i^{1+\delta}}\right]^{-1}.
\end{equation}
Define $\calI_i^K = \{\bx : G(\bx)=i\}$. Using the Shannon code for the distribution in \eqref{dist_for_ints}, for any distribution $Q^K$ on $\calX^K$, we have
\begin{align}
\bE_{Q^K} l(\bx) &= \sum_i Q^K(\calI_i^K)\ceil*{-\log p_i}\\
&\leq 1 - \sum_i Q^K(\calI_i^K) \log p_i\\
&= 1-\sum_i Q^K(\calI_i^K)\log c(\delta) + (1+\delta)\sum_i Q^K(\calI_i^K)\log i\\
&= 1-\log c(\delta) + (1+\delta)\sum_{\bx} Q^K(\bx) \log G(\bx).
\end{align}
However, since $\bE_{Q^K} l(\bx)$ is the expected length of a rate--distortion code for blocks of length $K$ of the source $Q^K$ at a (per--letter) distortion level $D$, we have,
\begin{align}
\sum_{\bx} Q^K(\bx) \log G(\bx) &\ge \frac{\bE_{Q^K} l(\bx)-1+\log c(\delta)}{1+\delta}\\
&\ge \frac{R_K(D,Q^K)-1+\log c(\delta)}{1+\delta}.
\end{align}
Hence, for any source $P$ with a marginal distribution $P^K$ and any $Q^K$, we have
\begin{align}
\bE_{P^K}\{G^\rho(\bX)\}&=\sum_{\bx \in \calX^K}P^K(\bx)G^\rho(\bx)\\
&= \sum_{\bx \in \calX^K}Q^K(\bx) \exp_2\left\{-\log\frac{Q^K(\bx)}{P^K(\bx)G^\rho(\bx)}\right\}\\
&\ge \exp_2\left\{-\sum_{\bx \in \calX^K} Q^K(\bx) \log\frac{Q^K(\bx)}{P^K(\bx)G^\rho(\bx)}\right\}\label{EPK-a} \\
&= \exp_2\left\{-D(Q^K\|P^K) + \rho\sum_{\bx \in \calX^K} Q^K(\bx) \log G(\bx)\right\}\\
& \ge \exp_2\left\{-D(Q^K\|P^K) + \rho\frac{R_K(D,Q^K)-1+\log c(\delta)}{1+\delta}\right\},
\end{align}
where \eqref{EPK-a} follows from Jensen's inequality. Since the above holds for any $Q^K$, we have, for any $\delta>0$,
\begin{equation}
\inf_{\calG_K}\frac{1}{K}\log \bE_{P^K}\{G^\rho(\bX)\} \ge \sup_{Q^K} \frac{1}{K} \left[ \frac{\rho}{1+\delta}R_K(D,Q^K) -D(Q^K\|P^K)  +O\left(\rho \frac{\log c(\delta)-1}{1+\delta}\right) \right].
\end{equation}
At this point, a remark on $\delta$ is in order. Note that $\delta$ is a parameter of a \emph{code for the integers}, regardless of $K$ and the guessing strategy used. $c(\delta)$ is a normalizing constant, with $c(\delta)\to\infty$ as $\delta \to 0$. In fact, $c(\delta) = \zeta(1+\delta)$, where $\zeta(\cdot)$ is the Riemann Zeta function, satisfying $\lim_{\delta \to 0} \delta \zeta(1+\delta) = 1$. Hence at the limit of $\delta \to \infty$, $\log c(\delta)$ can be approximated by $-\log \delta$. Yet, due to the independence in $K$, we have 
 \begin{equation}
\liminf_{K \to\infty}\inf_{\calG_K}\frac{1}{K}\log \bE_{P^K}\{G^\rho(\bX)\} \ge \liminf_{K \to\infty} \sup_{Q^K}\frac{1}{K}\left[\frac{\rho}{1+\delta}R_K(D,Q^K) -D(Q^K\|P^K)\right],
\end{equation}
for any $\delta>0$, which completes the proof. 
\end{proof}
\appendix
\section{Appendix - Additional Proofs and Technical Claims}\label{app proofs}
\subsection{Proof of \Cref{lemma for alternative form}}\label{proof for alternative form}
First, we derive an alternative expression to $\Gamma(Q_Y)$. We have
\begin{eqnarray}\label{alternative for Gamma}
	\Gamma(Q_Y)&=&\inf_{Q_{X|Y}}D(Q_{Y|X}\|W|Q_X)\\ 
	&=&\inf_{Q_{X|Y}}\sum_{x,y}Q_{XY}(x,y)\log\frac{Q_{Y|X}(y|x)}{W(y|x)}\\
	&=&\inf_{Q_{X|Y}}\sum_{x,y}Q_{XY}(x,y)\log\frac{Q_{X|Y}(x|y)Q_Y(y)}{Q_X(x)W(y|x)}\\
	&=&-H_Q(Y)+\inf_{Q_{X|Y}}\sum_{x,y}Q_{XY}(x,y)\log\frac{Q_{X|Y}(x|y)}{Q_X(x)W(y|x)}\\
	&=&-H_Q(Y)+\inf_V\inf_{Q_{X|Y}}\sum_{y}Q_{Y}(y)\sum_xQ_{X|Y}(x|y)\log\frac{Q_{X|Y}(x|y)}{V(x)W(y|x)}\label{alternative-GammaQY-a}\\
	&=&-H_Q(Y)+\inf_V\left\{-\sum_{y}Q_{Y}(y)\log\left[\sum_xV(x)W(y|x)\right]\right\}.\label{alternative-GammaQY-b}
\end{eqnarray}
In the above chain, \eqref{alternative-GammaQY-a} is true since
\begin{align}
\sum_{x,y}Q_{XY}(x,y)\log\frac{Q_{X|Y}(x|y)}{V(x)W(y|x)}-\sum_{x,y}Q_{XY}(x,y)&\log\frac{Q_{X|Y}(x|y)}{Q_X(x)W(y|x)} \\ &= \sum_{x,y}Q_{XY}(x,y) \log\frac{Q_X(x)}{V(x)} \ge 0
\end{align}
with equality when $V(x)=Q_X(x)$ for all $x$. \eqref{alternative-GammaQY-b} follows by noting that 
\begin{eqnarray}
\sum_xQ_{X|Y}(x|y)\log\frac{Q_{X|Y}(x|y)}{V(x)W(y|x)} &=& \sum_xQ_{X|Y}(x|y)\log\frac{[\sum_{\hat{x}}V(\hat{x})W(y|\hat{x}) ]Q_{X|Y}(x|y)}{ [\sum_{\hat{x}}V(\hat{x})W(y|\hat{x}) ] V(x)W(y|x)}\label{the divergence trick1}\\
&=& D\left(Q_{X|Y} || \tilde{V}_{X|Y}\right) - \log \left[\sum_{\hat{x}}V(\hat{x})W(y|\hat{x}) \right]\label{the divergence trick2},
\end{eqnarray}
where $\tilde{V}_{X|Y}(x|y)=\frac{1}{\sum_{\hat{x}}V(\hat{x})W(y|\hat{x})}V(x)W(y|x)$ and the expression is clearly minimized by choosing $Q_{X|Y} = \tilde{V}_{X|Y}$. Now,
\begin{align}
	\hat{R}_W&(D,Q_Y)
	\\
	&=\inf_{\{Q_{Y'|Y}:~\bE_Qd(Y,Y')\le D\}}[H_Q(Y')-H_Q(Y'|Y)+\Gamma(Q_{Y'})]\\
	&=\inf_{\{Q_{Y'|Y}:~\bE_Qd(Y,Y')\le D\}}\inf_V\left\{-H_Q(Y'|Y)-
	\sum_{y'}Q_{Y'}(y')\left[\log\sum_xV(x)W(y'|x)\right]\right\}\\
	&=\inf_{Q_{Y'|Y}}\sup_{s\ge 0}\inf_V\bigg[\sum_{y}Q_Y(y)\sum_{y'}Q_{Y'|Y}(y'|y)\log\frac{Q_{Y'|Y}(y'|y)}{
		\sum_xV(x)W(y'|x)}+\label{alternative-hatR-a}\\
	& \hspace{3cm} s\bigg(\sum_{y}Q_Y(y)\sum_{y'}Q_{Y'|Y}(y'|y)d(y,y')-D\bigg)\bigg]\\
	&=\inf_{Q_{Y'|Y}}\inf_V\sup_{s\ge 0}\sum_{y}Q_Y(y)
	\sum_{y'}Q_{Y'|Y}(y'|y)\bigg[\log\frac{Q_{Y'|Y}(y'|y)}{
		\sum_xV(x)W(y'|x)}+ sd(y,y')-sD\bigg]\\
	&=\inf_V\inf_{Q_{Y'|Y}}\sup_{s\ge 0}\sum_{y}Q_Y(y)
	\sum_{y'}Q_{Y'|Y}(y'|y)\left[\log\frac{Q_{Y'|Y}(y'|y)}{e^{-sd(y,y')}
		\sum_xV(x)W(y'|x)}-sD\right]\\
	&=\inf_V\sup_{s\ge 0}\inf_{Q_{Y'|Y}}\sum_{y}Q_Y(y)
	\sum_{y'}Q_{Y'|Y}(y'|y)\left[\log\frac{Q_{Y'|Y}(y'|y)}{e^{-sd(y,y')}
		\sum_xV(x)W(y'|x)}-sD\right]\label{alternative-hatR-b}\\
	&=\inf_V\sup_{s\ge 0}\left\{-\sum_{y}Q_Y(y)\log\left[
		\sum_{y'}e^{-sd(y,y')}
		\sum_xV(x)W(y'|x)\right]-sD\right\}\label{alternative-hatR-c}\\
	&=\inf_V\sup_{s\ge 0}\left\{-\sum_{y}Q_Y(y)\log\left[
		\sum_xV(x)U_s(x,y)\right]-sD\right\}\label{alternative-hatR-d}\\
	&=\sup_{s\ge 0}\inf_V\left\{-\sum_{y}Q_Y(y)\log\left[
		\sum_xV(x)U_s(x,y)\right]-sD\right\},
\end{align}
where \eqref{alternative-hatR-a} is due to the Lagrange multipliers analogous form for the optimization problem. \eqref{alternative-hatR-b} is since the utility function is convex in $Q_{Y'|Y}$ and concave (in fact, linear) in $s$. \eqref{alternative-hatR-c} uses a similar technique to \eqref{the divergence trick1}-\eqref{the divergence trick2}, namely, the expression 
\begin{equation}
\sum_{y'}Q_{Y'|Y}(y'|y)\log\frac{Q_{Y'|Y}(y'|y)}{e^{-sd(y,y')}\sum_xV(x)W(y'|x)}
\end{equation}
can be viewed as a divergence minus a normalization term. In \eqref{alternative-hatR-d} we have defined
\begin{equation}
	U_s(x,y)\dfn\sum_{y'}W(y'|x)e^{-sd(y,y')}
\end{equation}
and the last equality follows from the convexity in $V(x)$. Finally,
\begin{eqnarray}
	E(\rho)&=&\sup_{Q_Y}[\rho\hat{R}(D,Q_Y)-D(Q_Y\|P)]\\
	&=&\sup_{Q_Y}\sup_{s\ge 0}\inf_V\left\{\sum_yQ_Y(y)\log\frac{P(y)}
	{Q_Y(y)\left[\sum_xV(x)U_s(x,y)\right]^\rho}-\rho sD\right\}\\
	&=&\sup_{s\ge 0}\sup_{Q_Y}\inf_V\left\{\sum_yQ_Y(y)\log\frac{P(y)}
	{Q_Y(y)\left[\sum_xV(x)U_s(x,y)\right]^\rho}-\rho sD\right\}\\
	&=&\sup_{s\ge 0}\inf_V\sup_{Q_Y}\left\{\sum_yQ_Y(y)\log\frac{P(y)}
	{Q_Y(y)\left[\sum_xV(x)U_s(x,y)\right]^\rho}-\rho sD\right\}\label{Erho-a}\\
	&=&\sup_{s\ge 0}\inf_V\left\{\log\left(\sum_y\frac{P(y)}
	{\left[\sum_xV(x)U_s(x,y)\right]^\rho}\right)-\rho sD\right\}\label{Erho-b}\\
	&=&\sup_{s\ge 0}\inf_{M\in\calC\calH(W)}\left\{\log\left(\sum_y\frac{P(y)}
	{\left[\sum_{y'}M(y')e^{-sd(y,y')}\right]^\rho}\right)-\rho sD\right\}\label{Erho-c}\\
	&=&\sup_{s\ge 0}\inf_{M\in\calC\calH(W)}\left\{\log\left(\sum_y\frac{P(y)}
	{\left[\sum_{y'}M(y')e^{s[D-d(y,y')]}\right]^\rho}\right)\right\}.
\end{eqnarray}
In the above chain, \eqref{Erho-a} is since the utility is convex in $V(\cdot)$ and concave in $Q_Y(\cdot)$ and \eqref{Erho-b}, again, uses a similar technique to \eqref{the divergence trick1}-\eqref{the divergence trick2}. For \eqref{Erho-c}, remember that $U_s(x,y)\dfn\sum_{y'}W(y'|x)e^{-sd(y,y')}$ and define $M(y')=\sum_xV(x)W(y'|x)$. In fact, $M(y')$ is the channel output distribution
when its input is i.i.d.\ according to $V$. \eqref{Erho-c} thus replaces the minimization over $V$ by minimization over $M$, which is limited to 
$\calC\calH(W)$, that is, the convex hull of $\{W(\cdot|x),~x\in\calX\}$.
\subsection{Proof of \Cref{claim on sum over S}}
To ease notation, we write $P^n(\hat{\bx}|\bx) = \prod_{i=1}^{n}P(\hat{x}_i|x_i)$ as $P^n_{\hat{\bx}|\bx}$ in what follows. We have
\begin{align}
\sum_{\hat{\bx} \in \calS(\bx)}P^n(\hat{\bx}|\bx) &= P^n_{\hat{\bx}|\bx}\left( d(\bx,\hat{\bx}) \leq nD \right)
\\
&= P^n_{\hat{\bx}|\bx}\left( d(\bx,\hat{\bx}) -n\Dm \leq n\delta \right)
\\
&= P^n_{\hat{\bx}|\bx}\left( \sum_{i=1}^{n}d(x_i,\hat{x}_i) -n\Dm \leq n\delta \right)
\\
&= P^n_{\hat{\bx}|\bx}\left( \sum_{x\in \calX} \sum_{i: x_i=x}d(x,\hat{x}_i) -n\Dm \leq n\delta \right)
\\
&= P^n_{\hat{\bx}|\bx}\left( \sum_{x\in \calX} \frac{1}{n}\sum_{i: x_i=x}d(x,\hat{x}_i) -\Dm \leq \delta \right).
\end{align}
Since $P(\hat{x}|x)$ is the conditional distribution achieving $\Dm$, denoting 
\begin{equation}
\Dm(x) \dfn \sum_{\hat{x} \in \hat{\calX}} P(\hat{x}|x)d(x,\hat{x}),
\end{equation}
we have $\sum_{x \in \calX} \hat{P}_{\bx}(x) \Dm(x) = \Dm$. Thus, denoting by $N_{\bx}(x)$ the number of occurrences of $x$ in $\bx$, and analogously $N_{\bx \hat{\bx}}(x,\hat{x})$ the number of occurrences of the pair $(x,\hat{x})$, we have
\begin{align}
\sum_{\hat{\bx} \in \calS(\bx)}P^n(\hat{\bx}|\bx) &= P^n_{\hat{\bx}|\bx}\left( \sum_{x\in \calX} \left[ \frac{N_{\bx}(x)}{n}\sum_{\hat{x}\in \hat{\calX}} \frac{N_{\bx \hat{\bx}}(x,\hat{x})}{N_{\bx}(x)}d(x,\hat{x}_i) - \hat{P}_{\bx}(x)\Dm(x) \right] \leq \delta \right)
\\
&= P^n_{\hat{\bx}|\bx}\left( \sum_{x\in \calX} \hat{P}_{\bx}(x)\left[ \sum_{\hat{x}\in \hat{\calX}} \frac{N_{\bx \hat{\bx}}(x,\hat{x})}{N_{\bx}(x)}d(x,\hat{x}_i) - \Dm(x) \right] \leq \delta \right)
\\
&\ge P^n_{\hat{\bx}|\bx}\left( \forall_x \left[ \sum_{\hat{x}\in \hat{\calX}} \frac{N_{\bx \hat{\bx}}(x,\hat{x})}{N_{\bx}(x)}d(x,\hat{x}_i) - \Dm(x) \right] \leq \delta \right)
\\
&\ge P^n_{\hat{\bx}|\bx}\left( \forall_x \left| \frac{N_{\bx \hat{\bx}}(x,\hat{x})}{N_{\bx}(x)} - P(\hat{x}|x) \right| \leq \frac{\delta}{|\hat{\calX}|D_{max}} \right)
\\
&\ge 1-\epsilon'_n
\end{align}
for some $\epsilon'_n$ such that $\epsilon'_n \to 0$ as $n \to \infty$, where the last inequality is guaranteed by the law of large numbers, as each entry in $\hat{\bx}$ was drawn independently according to $P(\hat{x}|x)$. Note that since the law of large numbers is applied on independent drawings from $P(\hat{x}|x)$, $\epsilon'_n \to 0$ uniformly in $\bx$.
\subsection{Proof of \Cref{claim on sum over tc}}
Following the same notation as in the proof of \Cref{claim on sum over S}, we have
\begin{align}
\sum_{\hat{\bx} \in \calT^c(\bx)}P^n(\hat{\bx}|\bx) &= P^n_{\hat{\bx}|\bx} \left( \log\frac{P^n(\hat{\bx}|\bx)}{Q^n(\hat{\bx})} > n \left[ R(\Dm,\hat{P}_{\bx})+\epsilon\right] \right)
\\
&= P^n_{\hat{\bx}|\bx} \left( \frac{1}{n}\sum_{i=1}^{n}\log\frac{P(\hat{x}_i|x_i)}{Q(\hat{x}_i)} > R(\Dm,\hat{P}_{\bx})+\epsilon \right)
\\
&= P^n_{\hat{\bx}|\bx} \left( \frac{1}{n}\sum_{i=1}^{n}\log\frac{P(\hat{x}_i|x_i)}{Q(\hat{x}_i)} > \sum_{x\in \calX}\hat{P}_{\bx}(x)\sum_{\hat{x} \in \hat{\calX}}P(\hat{x}|x)\log\frac{P(\hat{x}|x)}{Q(\hat{x})} +\epsilon \right)
\\
&= P^n_{\hat{\bx}|\bx} \Bigg( \sum_{x\in \calX} \frac{N_{\bx}(x)}{n}\sum_{\hat{x}\in \hat{\calX}} \frac{N_{\bx \hat{\bx}}(x,\hat{x})}{N_{\bx}(x)}\log\frac{P(\hat{x}|x)}{Q(\hat{x})} 
\\
& \hspace{5cm} > \sum_{x\in \calX}\hat{P}_{\bx}(x)\sum_{\hat{x} \in \hat{\calX}}P(\hat{x}|x)\log\frac{P(\hat{x}|x)}{Q(\hat{x})} +\epsilon \Bigg)
\\
&= P^n_{\hat{\bx}|\bx} \left( \sum_{x\in \calX}\hat{P}_{\bx}(x) \sum_{\hat{x} \in \hat{\calX}} \left[ \frac{N_{\bx \hat{\bx}}(x,\hat{x})}{N_{\bx}(x)} - P(\hat{x}|x)\right] \log\frac{P(\hat{x}|x)}{Q(\hat{x})} > \epsilon \right) 
\\
&\leq P^n_{\hat{\bx}|\bx} \left( \sum_{x\in \calX}\hat{P}_{\bx}(x) \sum_{\hat{x} \in \hat{\calX}} \left[ \frac{N_{\bx \hat{\bx}}(x,\hat{x})}{N_{\bx}(x)} - P(\hat{x}|x)\right]  > \frac{\epsilon}{\log\frac{1}{\min_{\hat{x}}Q(\hat{x})}} \right)
\\
&\leq P^n_{\hat{\bx}|\bx} \left( \max_{x,\hat{x}} \left[ \frac{N_{\bx \hat{\bx}}(x,\hat{x})}{N_{\bx}(x)} - P(\hat{x}|x)\right]  > \frac{\epsilon}{|\hat{\calX}|\log\frac{1}{\min_{\hat{x}}Q(\hat{x})}} \right)
\\
&\leq \epsilon''_n,
\end{align}
where, again, the last inequality is due to the law of large numbers, $\frac{N_{\bx \hat{\bx}}(x,\hat{x})}{N_{\bx}(x)} \to P(\hat{x}|x)$ for all $x,\hat{x}$, since the alphabets $\calX,\hat{\calX}$ are finite and, without loss of generality, since $\min_{\hat{x}}Q(\hat{x})>0$. Again, the convergence is uniform in $\bx$.
\subsection{Proof of \Cref{claim ZIv's inequality}}
The proof is based on viewing any block-memoryless probability assignment for $\hat{\bx}$, followed by a Shannon code, as a finite-state encoder for $\hat{\bx}$. As such, its code length cannot be significantly shorter than the length of the code assigned by the LZ encoder \cite[Theorem 1]{ZL78}. It is similar to the line of argument used in \cite[eq. (32)]{Merhav20a}.

Let $M^{K+k}(\cdot)$ be any distribution on $\hat{\calX}^{K+k}$, and assume the sequence $\hat{\bx}$ is divided into blocks of length  $K+k$ and encoded using a Shannon code for $M^{K+k}(\cdot)$. The resulting compression ratio satisfies
\begin{align}
\rho_{K+k} &= \frac{1}{n \log|\hat{\calX}|}\sum_{i=1}^{m} \left\lceil -\log M^{K+k}\left({\hat{x}}_{(i-1)(K+k)+1}^{i(K+k)}\right) \right\rceil
\\
& \leq \frac{1}{n \log|\hat{\calX}|}\left[ -\sum_{i=1}^{m} \log M^{K+k}\left({\hat{x}}_{(i-1)(K+k)+1}^{i(K+k)}\right) +m \right]
\\
& = \frac{1}{n \log|\hat{\calX}|}\left[ - \log \prod_{i=1}^{m}M^{K+k}\left({\hat{x}}_{(i-1)(K+k)+1}^{i(K+k)}\right) +m \right].
\end{align}
The above encoder can clearly be implemented using a finite-state machine of $s(K+k)$ states, where $s(\cdot)$ is independent of $n$.  Hence, by \cite[Theorem 1]{ZL78}, dropping the dependence on $K+k$ for ease of notation, we have
\begin{equation}
\rho_{K+k} \ge \frac{c(\hat{\bx}) +s^2}{n \log|\hat{\calX}|} \log\left(\frac{c(\hat{\bx}) +s^2}{4s^2}\right) + \frac{2s^2}{n \log|\hat{\calX}|}.
\end{equation}
Consequently, 
\begin{align}
- \log \prod_{i=1}^{m}M^{K+k}\left({\hat{x}}_{(i-1)(K+k)+1}^{i(K+k)}\right) +m &\ge (c(\hat{\bx}) +s^2) \log\left(\frac{c(\hat{\bx}) +s^2}{4s^2}\right) + 2s^2
\\
& \ge (c(\hat{\bx}) +1 +s^2 -1) \log\left(\frac{2|\hat{\calX}|}{2|\hat{\calX}|}\frac{c(\hat{\bx}) +1}{4s^2}\right) + 2s^2
\\
& = (c(\hat{\bx}) +1)\log\left(2|\hat{\calX}| (c(\hat{\bx}) +1)\right)
\\
& \qquad + (s^2-1)\log\left(2|\hat{\calX}| (c(\hat{\bx}) +1)\right)
\\
& \qquad + 2s^2 - (c(\hat{\bx}) +s^2)\log\left(8|\hat{\calX}| s^2\right).
\end{align}
Finally, by \cite[Theorem 2]{ZL78}, $LZ(\hat{\bx}) \leq (c(\hat{\bx}) +1)\log\left(2|\hat{\calX}| (c(\hat{\bx}) +1)\right)$, thus
\begin{equation}
LZ(\hat{\bx}) \leq -\log \prod_{i=1}^{m}M^{K+k}\left({\hat{x}}_{(i-1)(K+k)+1}^{i(K+k)}\right) + m + (c(\hat{\bx}) +s^2)\log\left(8|\hat{\calX}| s^2\right).
\end{equation}
Setting $\nu_{K+k}(n) = m + (c(\hat{\bx}) +s^2)\log\left(8|\hat{\calX}| s^2\right)$, $m=\frac{n}{K+k}$ and since $c(\hat{\bx}) \leq \frac{n \log |\hat{\calX}|}{(1-\epsilon_n) \log n}$ (\cite[Theorem 2]{LZ76}), we obtain the required result.
\subsection{Proof of \Cref{probability of gap oblovious}}\label{appendix proof of probability of gap oblovious}
Since $Q$ is $\Psi$-mixing, for any two block $\ux_i, \ux_{i+1}$ which are $k$ symbols apart, we have
\begin{equation}
\left| \frac{Q(\ux_i,\ux_{i+1})}{Q(\ux_i)Q(\ux_{i+1})} - 1\right| \leq \epsilon_k
\end{equation}
for some $\epsilon_k \to 0$ as $k \to \infty$. Hence,
\begin{align}
Q\left(G_{\bx}^K\right) &= \sum_{\tilde{\bx} \in \calX^n : \tilde{\ux}_i = \ux_i , 1 \leq i \leq m}Q(\tilde{\bx})
\\
&= Q(\ux_1, \ldots, \ux_{m})
\\
& \leq (1+\epsilon_k)^{m-1} \prod_{i=1}^{m}Q^K(\ux_i) 
\\
& = (1+\epsilon_k)^{m-1} 2^{-m \left(D\left(\hat{P}^K_{\bx}||Q^K\right) + \hat{H}_{\bx}^K(X^K)\right)}.
\end{align}
The opposite inequality follows in a similar manner. 
\subsection{Proof of \Cref{size of generalized type}}\label{appendix proof of size of generalized type}
Let $P$ be any stationary $\Psi$-mixing distribution with a $K$-th order marginal $P^K$. Then,
\begin{align}
1 &\ge P\left(T^K(P^K)\right)
\\ 
& = \sum_{G_{\bx}^K \in T^K(P^K)} P^K\left(G_{\bx}^K\right)
\\
& \ge \left| T^K(P^K) \right|_G (1-\epsilon_k)^{m-1} 2^{-m\hat{H}_{\bx}^K(X^K)}.
\end{align}
\subsection{Converse and Direct for Individual Sequences and Finite--State Guessing}\label{individual sequences}
\subsubsection{Converse}
The converse result is similar to the converse in \cite{Merhav20a}, with the exception of a different notion of a \emph{success probability}. Specifically, for any sequence $\bx$, we first generalize the success probability of the finite--state machine, $P(\bx)$ in the notation of \cite{Merhav20a} when one has to guess the exact sequence $\bx$, to a success probability in terms of guessing under a distortion constraint. That is, by \cite[eq.\ (21)]{Merhav20a}, for any finite--state guessing machine $F$, which is fed by i.i.d.\ random bits, the $\rho$--th moment satisfies the following lower bound
\begin{equation}
    \bE\{G_F^\rho|\bx\} \ge \frac{2^{-\rho}}{e^2}\exp_2\left\{-\rho \log P(\bx)\right\},
\end{equation}
where $P(\bx)$ is the probability that the machine output is $\bx$. Note that similar to \Cref{geometric lemma}, the bound above is based on bounding the $\rho$--th moment of a Geometric random variable, i.e., repeated independent trials, where success is declared when the machine outputs the correct sequence. Thus, analogously, when guessing subject to a distortion measure $d$ and value $D$, each trail is independent as well (as the machine is restarted after each guess), yet success is declared when the machine outputs a sequence within the designated distortion. Hence, we have 
\begin{equation}
    \bE\{G_F^\rho|\bx\} \ge \frac{2^{-\rho}}{e^2}\exp_2\left\{-\rho \log P[\calS(\bx)]\right\},
\end{equation}
where $\calS(\bx)$, the success set, was defined below \eqref{distortion measure}. We thus have
\begin{equation}\label{bound on E using P (individual)}
    \bE\{G_F^\rho|\bx\} \ge \frac{2^{-\rho}}{e^2}\exp_2\left\{-\rho \log  \sum_{\{\hat{\bx}:~d(\bx,\hat{\bx})\le nD\}} P(\hat{\bx})\right\}.
\end{equation}
$P(\hat{\bx})$, the probability that the machine outputs $\hat{\bx}$, can be viewed as a probability assignment on the sequences $\hat{\bx} \in \hat{\calX}^n$. As such, $-\log P(\hat{\bx})$ cannot be too small compared to the code length of an LZ encoder. This is similar to the argument behind \Cref{claim ZIv's inequality} used on a block--memoryless distribution. In the context of the current discussion, however, it is easier to harness \cite[equations (31) and (32)]{Merhav20a}, resulting in
\begin{equation}\label{bound on P using clogc}
-\log P(\hat{\bx}) \ge c(\hat{\bx})\log c(\hat{\bx}) -\frac{n \log[4K^2(l)]\log|\hat{\calX}|}{(1-\epsilon_n)\log n} -K^2(l) \log[4K^2(l)] -m\log(2s^3 e),
\end{equation}
where $s$ is the number of states, $m$ and $l$ are integers such that $ml=n$ and $K(l)$ is the number of states of a machine implementing a Shannon code on blocks of size $l$. Thus, choosing $l$ to grow to infinity slow enough, we have $-\log P(\hat{\bx}) \gexe LZ(\hat{\bx})$. Specifically, with $l=\log(\log n)$ and since $K(l)$ is bounded by $l \alpha^l$, we have
\begin{equation}
\frac{n \log[4K^2(l)]\log|\hat{\calX}|}{(1-\epsilon_n)\log n} +K^2(l) \log[4K^2(l)] +m\log(2s^3 e) = o(n)
\end{equation}
and hence
\begin{eqnarray}
    \bE\{G_F^\rho|\bx\} &\gexe& \exp_2\left\{-\rho \log \sum_{\{\hat{\bx}:~d(\bx,\hat{\bx})\le nD\}} 2^{-LZ(\hat{\bx})}\right\} \\
    &=& \left[\sum_{\{\hat{\bx}:~d(\bx,\hat{\bx})\le nD\}} 2^{-LZ(\hat{\bx})}\right]^{-\rho},
\end{eqnarray}
which proves part (i) of the converse result.

For part (ii), instead of \eqref{bound on P using clogc}, we lower bound $-\log P(\hat{\bx})$ using the finite--state compressability of $\hat{\bx}$, that is, by \cite[eq. (31)]{Merhav20a},
\begin{equation}
-\log P(\hat{\bx}) \ge n \rho_{K(l)}(\hat{\bx}) -m\log(2 s^3 e).
\end{equation}
Continuing from \eqref{bound on E using P (individual)}, we have
\begin{eqnarray}
    \bE\{G_F^\rho|\bx\} &\ge& \frac{2^{-\rho}}{e^2}\exp_2\left\{-\rho \log  \sum_{\{\hat{\bx}:~d(\bx,\hat{\bx})\le nD\}} \exp_2\left\{-n \rho_{K(l)}(\hat{\bx}) +m\log\left(2 s^3 e\right) \right\} \right\}\\
    &=&\frac{2^{-\rho}}{e^2}\left[  \sum_{\{\hat{\bx}:~d(\bx,\hat{\bx})\le nD\}} \exp_2\left\{-n \rho_{K(l)}(\hat{\bx}) +m\log\left(2 s^3 e\right) \right\} \right]^{-\rho},
\end{eqnarray}
which is true for any integers $m$ and $l$ such that $ml=n$. This completes the proof of the converse.
\subsubsection{Direct}
For the direct, we first note that the universal distribution given in \eqref{tP with LZ}, that is $\tP(\hat{\bx})\exe 2^{-LZ(\hat{\bx})}$, results in
\begin{equation}
	\tP[\calS(\bx)]\exe \sum_{\{\hat{\bx}:~d(\bx,\hat{\bx})\le nD\}}2^{-LZ(\hat{\bx})}.
\end{equation}
Thus, invoking \Cref{geometric lemma} yields
\begin{equation}\label{direct for individual}
	\bE\{G_{LZ}^\rho|\bx\}\exe \left[\sum_{\{\hat{\bx}:~d(\bx,\hat{\bx})\le nD\}} 2^{-LZ(\hat{\bx})}\right]^{-\rho}
\end{equation}
for the specific universal distribution in \eqref{tP with LZ}. Clearly, this meets the converse result. However, the distribution in \eqref{tP with LZ} cannot be implemented with a finite--state machine whose number of states is independent of $n$. To asymptotically achieve \eqref{direct for individual} with such a finite--state machine, we invoke the solution in \cite{Merhav20a} for the same problem (as the same distribution was sought). That is, we consider the distribution in \eqref{block-LZ}. This distribution can be implemented with a finite--state machine with $s(l)$ states, and for such a finite--state guessing machine we have
\begin{eqnarray}
\bE\{G_{LZ}^\rho|\bx\} &\exe& \left[\sum_{\{\hat{\bx}:~d(\bx,\hat{\bx})\le nD\}} \prod_{i=0}^{n/l-1}\left[\frac{2^{-LZ({\hat{x}}_{il+1}^{il+l})}}{\sum_{\hat{x}^l \in \hat{\calX}^l}2^{-LZ(\hat{x}^l)}} \right]\right]^{-\rho} \\
&\exe& \left[\sum_{\{\hat{\bx}:~d(\bx,\hat{\bx})\le nD\}} \prod_{i=0}^{n/l-1} 2^{-LZ({\hat{x}}_{il+1}^{il+l})} \right]^{-\rho} \\
&=& \left[\sum_{\{\hat{\bx}:~d(\bx,\hat{\bx})\le nD\}} \exp_2\left\{ -\sum_{i=0}^{n/l-1} LZ({\hat{x}}_{il+1}^{il+l}) \right\}\right]^{-\rho} \\
&\leq& \left[\sum_{\{\hat{\bx}:~d(\bx,\hat{\bx})\le nD\}} \exp_2\left\{ -\sum_{i=0}^{n/l-1} c({\hat{x}}_{il+1}^{il+l})\log c({\hat{x}}_{il+1}^{il+l}) - n \epsilon(l) \right\}\right]^{-\rho}\label{use of 33 from Neri's paper}
\\
&\leq& \Bigg[\sum_{\{\hat{\bx}:~d(\bx,\hat{\bx})\le nD\}} \exp_2\bigg\{ - n\Big[ \rho_{K}(\hat{\bx}) + \nonumber
\\
&& \hspace{4cm}\frac{\log(4K^2)}{(1-\epsilon(l))\log l}+ \frac{K^2\log(4K^2)}{l} +\epsilon(l)\Big]  \bigg\}\Bigg]^{-\rho},\label{use of 38 from Neri's paper}
\end{eqnarray}
where $\epsilon(l) \to 0$ as $l \to \infty$, \eqref{use of 33 from Neri's paper} is by \cite[eq. (33)]{Merhav20a} and \eqref{use of 38 from Neri's paper} is by \cite[eq. (38)]{Merhav20a}.
\bibliographystyle{IEEEtran}
\bibliography{guesswork}

\begin{thebibliography}{10}
\providecommand{\url}[1]{#1}
\csname url@samestyle\endcsname
\providecommand{\newblock}{\relax}
\providecommand{\bibinfo}[2]{#2}
\providecommand{\BIBentrySTDinterwordspacing}{\spaceskip=0pt\relax}
\providecommand{\BIBentryALTinterwordstretchfactor}{4}
\providecommand{\BIBentryALTinterwordspacing}{\spaceskip=\fontdimen2\font plus
\BIBentryALTinterwordstretchfactor\fontdimen3\font minus
  \fontdimen4\font\relax}
\providecommand{\BIBforeignlanguage}[2]{{%
\expandafter\ifx\csname l@#1\endcsname\relax
\typeout{** WARNING: IEEEtran.bst: No hyphenation pattern has been}%
\typeout{** loaded for the language `#1'. Using the pattern for}%
\typeout{** the default language instead.}%
\else
\language=\csname l@#1\endcsname
\fi
#2}}
\providecommand{\BIBdecl}{\relax}
\BIBdecl

\bibitem{wozencraft1957sequential}
J.~M. Wozencraft, ``Sequential decoding for reliable communication,'' Ph.D.
  dissertation, Research Laboratory of Electronics, Massachusetts Institute of
  Technology, May 1957.

\bibitem{Arikan96}
E.~Arikan, ``An inequality on guessing and its application to sequential
  decoding,'' \emph{IEEE Transactions on Information Theory}, vol.~42, no.~1,
  pp. 99--105, January 1996.

\bibitem{PfisterSullivan}
C.~E. Pfister and W.~G. Sullivan, ``R\'{e}nyi entropy, guesswork moments, and
  large deviations,'' \emph{IEEE Transactions on Information Theory}, vol.~50,
  no.~11, pp. 2794--2800, November 2004.

\bibitem{AM98joint}
E.~Arikan and N.~Merhav, ``Joint source-channel coding and guessing with
  application to sequential decoding,'' \emph{IEEE Transactions on Information
  Theory}, vol.~44, no.~5, pp. 1756--1769, 1998.

\bibitem{Szpankowski11F2V}
W.~Szpankowski and S.~Verd\'u, ``Minimum expected length of fixed-to-variable
  lossless compression without prefix constraints,'' \emph{IEEE Transactions on
  Information Theory}, vol.~57, no.~7, pp. 4017--4025, July 2011.

\bibitem{Kontoyiannis14optimal_lossless}
I.~Kontoyiannis and S.~Verd\'u, ``Optimal lossless data compression:
  Non-asymptotics and asymptotics,'' \emph{IEEE Transactions on Information
  Theory}, vol.~60, no.~2, pp. 777--795, February 2014.

\bibitem{Courtade_Verdu_14}
T.~A. {Courtade} and S.~{Verd\'u}, ``Cumulant generating function of codeword
  lengths in optimal lossless compression,'' in \emph{2014 IEEE International
  Symposium on Information Theory}, Honolulu, HI, USA, June 2014, pp.
  2494--2498.

\bibitem{Kosut_universal_F2V_17}
O.~Kosut and L.~Sankar, ``Asymptotics and non-asymptotics for universal
  fixed-to-variable source coding,'' \emph{IEEE Transactions on Information
  Theory}, vol.~63, no.~6, pp. 3757--3772, June 2017.

\bibitem{kumar2020guessing}
M.~A. Kumar, A.~Sunny, A.~Thakre, A.~Kumar, and G.~D. Manohar, ``Are guessing,
  source coding, and tasks partitioning birds of a feather?'' \emph{arXiv
  preprint arXiv:2012.13707}, 2020.

\bibitem{Bunte2014Tasks}
C.~Bunte and A.~Lapidoth, ``Encoding tasks and {R}\'enyi entropy,'' \emph{IEEE
  Transactions on Information Theory}, vol.~60, no.~9, pp. 5065--5076, 2014.

\bibitem{AM98}
E.~Arikan and N.~Merhav, ``Guessing subject to distortion,'' \emph{IEEE
  Transactions on Information Theory}, vol.~44, no.~3, pp. 1041--1056, May
  1998.

\bibitem{Duffy2019Capacity}
K.~R. Duffy, J.~Li, and M.~M\'{e}dard, ``Capacity-achieving guessing random
  additive noise decoding,'' \emph{IEEE Transactions on Information Theory},
  vol.~65, no.~7, pp. 4023--4040, 2019.

\bibitem{bishop1995improving}
M.~Bishop and D.~V. Klein, ``Improving system security via proactive password
  checking,'' \emph{Computers \& Security}, vol.~14, no.~3, pp. 233--249, 1995.

\bibitem{dell2010password}
M.~D. Amico, P.~Michiardi, and Y.~Roudier, ``Password strength: An empirical
  analysis,'' in \emph{Proceedings of IEEE Infocom}, San Diego, CA, USA, March
  2010, pp. 1--9.

\bibitem{kelley2012guess}
P.~G. Kelley, S.~Komanduri, M.~L. Mazurek, R.~Shay, T.~Vidas, L.~Bauer,
  N.~Christin, L.~F. Cranor, and J.~Lopez, ``Guess again (and again and again):
  Measuring password strength by simulating password-cracking algorithms,'' in
  \emph{2012 IEEE Symposium on Security and Privacy}, San Francisco, CA, USA,
  May 2012, pp. 523--537.

\bibitem{komanduri2011passwords}
S.~Komanduri, R.~Shay, P.~G. Kelley, M.~L. Mazurek, L.~Bauer, N.~Christin,
  L.~F. Cranor, and S.~Egelman, ``Of passwords and people: Measuring the effect
  of password-composition policies,'' in \emph{Proceedings of the SIGCHI
  Conference on Human Factors in Computing Systems}.\hskip 1em plus 0.5em minus
  0.4em\relax New York, NY, USA: ACM, 2011, pp. 2595--2604.

\bibitem{MerhavCohen20}
N.~{Merhav} and A.~{Cohen}, ``Universal randomized guessing with application to
  asynchronous decentralized brute--force attacks,'' \emph{IEEE Transactions on
  Information Theory}, vol.~66, no.~1, pp. 114--129, 2020.

\bibitem{malone2012investigating}
D.~Malone and K.~Maher, ``Investigating the distribution of password choices,''
  in \emph{Proceedings of the 21st international conference on World Wide
  Web}.\hskip 1em plus 0.5em minus 0.4em\relax Lyon, France: ACM, 2012, pp.
  301--310.

\bibitem{1341406}
J.~Yan, A.~Blackwell, R.~Anderson, and A.~Grant, ``Password memorability and
  security: empirical results,'' \emph{IEEE Security Privacy}, vol.~2, no.~5,
  pp. 25--31, September 2004.

\bibitem{Vishwakarma14dictionary}
D.~Vishwakarma and C.~E.~V. Madhavan, ``Efficient dictionary for salted
  password analysis,'' in \emph{2014 IEEE International Conference on
  Electronics, Computing and Communication Technologies (CONECCT)}, Bangalore,
  India, January 2014, pp. 1--6.

\bibitem{8590812}
Z.~Rui and Z.~Yan, ``A survey on biometric authentication: Toward secure and
  privacy-preserving identification,'' \emph{IEEE Access}, vol.~7, pp.
  5994--6009, 2019.

\bibitem{page2015ECG}
A.~Page, A.~Kulkarni, and T.~Mohsenin, ``Utilizing deep neural nets for an
  embedded ecg-based biometric authentication system,'' in \emph{2015 IEEE
  Biomedical Circuits and Systems Conference (BioCAS)}, Atlanta, GA, USA, 2015,
  pp. 1--4.

\bibitem{wiki_QRS}
\BIBentryALTinterwordspacing
W.~contributors, ``Qrs complex,'' online; accessed 29-September-2021. [Online].
  Available: \url{https://en.wikipedia.org/wiki/QRS\_complex}
\BIBentrySTDinterwordspacing

\bibitem{hammad2019ECG}
M.~Hammad, Y.~Liu, and K.~Wang, ``Multimodal biometric authentication systems
  using convolution neural network based on different level fusion of ecg and
  fingerprint,'' \emph{IEEE Access}, vol.~7, pp. 26\,527--26\,542, 2019.

\bibitem{aziz2019ecg}
S.~Aziz, M.~U. Khan, Z.~A. Choudhry, A.~Aymin, and A.~Usman, ``{ECG}-based
  biometric authentication using empirical mode decomposition and support
  vector machines,'' in \emph{2019 IEEE 10th Annual Information Technology,
  Electronics and Mobile Communication Conference (IEMCON)}.\hskip 1em plus
  0.5em minus 0.4em\relax Vancouver, BC, Canada: IEEE, 2019, pp. 0906--0912.

\bibitem{Boles2017VoiceBiometrics}
A.~Boles and P.~Rad, ``Voice biometrics: Deep learning-based voiceprint
  authentication system,'' in \emph{2017 12th System of Systems Engineering
  Conference (SoSE)}, Waikoloa, HI, USA, 2017, pp. 1--6.

\bibitem{meng2020active}
Z.~Meng, M.~U.~B. Altaf, and B.-H.~F. Juang, ``Active voice authentication,''
  \emph{Digital Signal Processing}, vol. 101, p. 102672, 2020.

\bibitem{mahfouz2017survey}
A.~Mahfouz, T.~M. Mahmoud, and A.~S. Eldin, ``A survey on behavioral biometric
  authentication on smartphones,'' \emph{Journal of information security and
  applications}, vol.~37, pp. 28--37, 2017.

\bibitem{gu2014towards}
S.~Gu and L.~Rigazio, ``Towards deep neural network architectures robust to
  adversarial examples,'' \emph{arXiv preprint arXiv:1412.5068}, 2014.

\bibitem{carlini2017provably}
N.~Carlini, G.~Katz, C.~Barrett, and D.~L. Dill, ``Provably minimally-distorted
  adversarial examples,'' \emph{arXiv preprint arXiv:1709.10207}, 2017.

\bibitem{bi2005support}
J.~Bi and T.~Zhang, ``Support vector classification with input data
  uncertainty,'' \emph{Advances in neural information processing systems},
  vol.~17, no.~1, pp. 161--168, 2005.

\bibitem{Tsai2021AdversarialRobustness}
Y.-L. Tsai, C.-Y. Hsu, C.-M. Yu, and P.-Y. Chen, ``Non-singular adversarial
  robustness of neural networks,'' in \emph{2021 IEEE International Conference
  on Acoustics, Speech and Signal Processing (ICASSP)}, Toronto, ON, Canada,
  2021, pp. 3840--3844.

\bibitem{xu2017feature}
W.~Xu, D.~Evans, and Y.~Qi, ``Feature squeezing: Detecting adversarial examples
  in deep neural networks,'' \emph{arXiv preprint arXiv:1704.01155}, 2017.

\bibitem{ZL78}
J.~Ziv and A.~Lempel, ``Compression of individual sequences via variable-rate
  coding,'' \emph{IEEE Transactions on Information Theory}, vol.~24, no.~5, pp.
  530--536, September 1978.

\bibitem{Massey94}
J.~L. Massey, ``Guessing and entropy,'' in \emph{Proceedings of IEEE
  International Symposium on Information Theory}, Trondheim, Norway, 1994, p.
  204.

\bibitem{kuzuoka2020asynchronous}
S.~Kuzuoka, ``Asynchronous guessing subject to distortion,'' \emph{arXiv
  preprint arXiv:2012.04901}, 2020.

\bibitem{kuzuoka2021asynchronous}
------, ``Asynchronous guessing subject to distortion,'' in \emph{Proceedings\
  2021 IEEE International Symposium on Information Theory}, Melbourne,
  Australia, July 2021, pp. 2008--2012.

\bibitem{SHBCM19}
S.~Salamatian, W.~Huleihel, A.~Beirami, A.~Cohen, and M.~M\'{e}dard, ``Why
  botnets work: Distributed brute-force attacks need no synchronization,''
  \emph{IEEE Transactions on Information Forensics and Security}, vol.~14,
  no.~9, pp. 2288--2299, 2019.

\bibitem{Weinberger_universal_ordering_92}
M.~J. Weinberger, J.~Ziv, and A.~Lempel, ``On the optimal asymptotic
  performance of universal ordering and of discrimination of individual
  sequences,'' \emph{IEEE Transactions on Information Theory}, vol.~38, no.~2,
  pp. 380--385, March 1992.

\bibitem{Beirami15geometric}
A.~Beirami, R.~Calderbank, M.~Christiansen, K.~Duffy, A.~Makhdoumi, and
  M.~M\'{e}dard, ``A geometric perspective on guesswork,'' in \emph{2005 53rd
  Annual Allerton Conference on Communication, Control, and Computing},
  Monticello, IL, USA, September 2015, pp. 941--948.

\bibitem{owens2008study}
J.~Owens and J.~Matthews, ``A study of passwords and methods used in
  brute-force {SSH} attacks,'' in \emph{USENIX Workshop on Large-Scale Exploits
  and Emergent Threats (LEET)}, San Francisco, CA, USA, 2008.

\bibitem{tirado2018new}
E.~Tirado, B.~Turpin, C.~Beltz, P.~Roshon, R.~Judge, and K.~Gagneja, ``A new
  distributed brute-force password cracking technique,'' in \emph{International
  Conference on Future Network Systems and Security}, Paris, France, 2018, pp.
  117--127.

\bibitem{SHBCM20}
S.~Salamatian, W.~Huleihel, A.~Beirami, A.~Cohen, and M.~M\'{e}dard,
  ``Centralized vs decentralized targeted brute-force attacks: Guessing with
  side-information,'' \emph{IEEE Transactions on Information Forensics and
  Security}, vol.~15, pp. 3749--3759, 2020.

\bibitem{sundaresan2007guessing}
R.~Sundaresan, ``Guessing under source uncertainty,'' \emph{IEEE Transactions
  on Information Theory}, vol.~53, no.~1, pp. 269--287, January 2007.

\bibitem{malone2004guesswork}
D.~Malone and W.~G. Sullivan, ``Guesswork and entropy,'' \emph{IEEE
  Transactions on Information Theory}, vol.~50, no.~3, pp. 525--526, March
  2004.

\bibitem{5673955}
M.~K. Hanawal and R.~Sundaresan, ``Guessing revisited: A large deviations
  approach,'' \emph{IEEE Transactions on Information Theory}, vol.~57, no.~1,
  pp. 70--78, January 2011.

\bibitem{christiansen2013guesswork}
M.~M. Christiansen and K.~R. Duffy, ``Guesswork, large deviations, and shannon
  entropy,'' \emph{IEEE Transactions Information Theory}, vol.~59, no.~2, pp.
  796--802, February 2013.

\bibitem{Merhav20a}
N.~{Merhav}, ``Guessing individual sequences: generating randomized guesses
  using finite--state machines,'' \emph{IEEE Transactions on Information
  Theory}, vol.~66, no.~5, pp. 2912--2920, 2020.

\bibitem{bonneau2012science}
J.~Bonneau, ``The science of guessing: Analyzing an anonymized corpus of 70
  million passwords,'' in \emph{2012 IEEE Symposium on Security and Privacy},
  San Francisco, CA, USA, May 2012, pp. 538--552.

\bibitem{christiansen2013guessing}
M.~M. Christiansen, K.~R. Duffy, F.~du~Pin~Calmon, and M.~M\'edard, ``Guessing
  a password over a wireless channel (on the effect of noise non-uniformity),''
  in \emph{2013 Asilomar Conference on Signals, Systems and Computers}, Pacific
  Grove, CA, USA, November 2013, pp. 51--55.

\bibitem{Merhav20b}
N.~{Merhav}, ``Noisy guesses,'' \emph{IEEE Transactions on Information Theory},
  vol.~66, no.~8, pp. 4796--4803, 2020.

\bibitem{christiansen2015multi}
M.~M. Christiansen, K.~R. Duffy, F.~du~Pin~Calmon, and M.~M{\'e}dard,
  ``Multi-user guesswork and brute force security,'' \emph{IEEE Transactions on
  Information Theory}, vol.~61, no.~12, pp. 6876--6886, December 2015.

\bibitem{7282958}
A.~Beirami, R.~Calderbank, K.~Duffy, and M.~M\'edard, ``Quantifying
  computational security subject to source constraints, guesswork and
  inscrutability,'' in \emph{2015 IEEE International Symposium on Information
  Theory (ISIT)}, Hong Kong, China, June 2015, pp. 2757--2761.

\bibitem{courtade14}
T.~A. Courtade and S.~Verd\'u, ``Variable-length lossy compression and channel
  coding: Non-asymptotic converses via cumulant generating functions,'' in
  \emph{2014 IEEE International Symposium on Information Theory}, Honolulu, HI,
  USA, June 2014, pp. 2499--2503.

\bibitem{sason2018tight}
I.~Sason, ``Tight bounds on the {R}{\'e}nyi entropy via majorization with
  applications to guessing and compression,'' \emph{Entropy}, vol.~20, no.~12,
  p. 896, 2018.

\bibitem{4036408}
R.~Sundaresan, ``Guessing under source uncertainty with side information,'' in
  \emph{2006 IEEE International Symposium on Information Theory}, Seattle, WA,
  USA, July 2006, pp. 2438--2440.

\bibitem{Sason_Verdu_18}
I.~{Sason} and S.~{Verd\'u}, ``Improved bounds on lossless source coding and
  guessing moments via {R}\'enyi measures,'' \emph{IEEE Transactions on
  Information Theory}, vol.~64, no.~6, pp. 4323--4346, June 2018.

\bibitem{yona17bias}
Y.~Yona and S.~Diggavi, ``The effect of bias on the guesswork of hash
  functions,'' in \emph{2017 IEEE International Symposium on Information Theory
  (ISIT)}, Aachen, Germany, June 2017, pp. 2248--2252.

\bibitem{Ardimanov20Oracle}
N.~Ardimanov, O.~Shayevitz, and I.~Tamo, ``Minimum guesswork with an unreliable
  oracle,'' \emph{IEEE Transactions on Information Theory}, vol.~66, no.~12,
  pp. 7528--7538, 2020.

\bibitem{weinberger2020guessing}
N.~Weinberger and O.~Shayevitz, ``Guessing with a bit of help,''
  \emph{Entropy}, vol.~22, no.~1, p.~39, 2020.

\bibitem{CT06}
T.~M. Cover and J.~A. Thomas, \emph{Elements of Information Theory},
  2nd~ed.\hskip 1em plus 0.5em minus 0.4em\relax New York, NY, USA: John Wiley
  \& Sons, 2006.

\bibitem{LZ76}
A.~Lempel and J.~Ziv, ``On the complexity of finite sequences,'' \emph{IEEE
  Transactions on information theory}, vol.~22, no.~1, pp. 75--81, 1976.

\bibitem{Feder91Gambling}
M.~Feder, ``Gambling using a finite state machine,'' \emph{IEEE Transactions on
  Information Theory}, vol.~37, no.~5, pp. 1459--1465, 1991.

\bibitem{FMG92}
M.~Feder, N.~Merhav, and M.~Gutman, ``Universal prediction of individual
  sequences,'' \emph{IEEE Transactions on Information Theory}, vol.~38, no.~4,
  pp. 1258--1270, July 1992.

\bibitem{Weissman2001Twofold}
T.~Weissman, N.~Merhav, and A.~Somekh-Baruch, ``Twofold universal prediction
  schemes for achieving the finite-state predictability of a noisy individual
  binary sequence,'' \emph{IEEE Transactions on Information Theory}, vol.~47,
  no.~5, pp. 1849--1866, 2001.

\bibitem{Modha2004FS-RD}
D.~Modha and D.~de~Farias, ``Finite-state rate-distortion for individual
  sequences,'' in \emph{Proceedings of the IEEE International Symposium on
  Information Theory}, Chicago, IL, USA, 2004, pp. 562--.

\bibitem{DUDE2005}
T.~Weissman, E.~Ordentlich, G.~Seroussi, S.~Verdu, and M.~Weinberger,
  ``Universal discrete denoising: known channel,'' \emph{IEEE Transactions on
  Information Theory}, vol.~51, no.~1, pp. 5--28, 2005.

\bibitem{Merhav12Encryption}
N.~Merhav, ``Perfectly secure encryption of individual sequences,'' \emph{IEEE
  Transactions on Information Theory}, vol.~59, no.~3, pp. 1302--1310, 2013.

\bibitem{berger1971rate}
T.~Berger, \emph{Rate Distortion Theory: A Mathematical Basis for Data
  Compression}, ser. Prentice-Hall electrical engineering series.\hskip 1em
  plus 0.5em minus 0.4em\relax Prentice-Hall, 1971.

\end{thebibliography}
\end{document}